\documentclass[pra,reprint,superscriptaddress,bibnotes]{revtex4-1}  

\usepackage[T1]{fontenc}
\usepackage[utf8]{inputenc} 
\usepackage[australian]{babel}
\usepackage{lmodern}

\usepackage[a4paper,centering,hmargin=1.75cm,vmargin=2cm]{geometry} 

\usepackage{amsmath,amssymb,graphicx,bm,microtype}
\usepackage[dvipsnames]{xcolor}
\usepackage{booktabs}
\usepackage{siunitx} 
\usepackage{braket}
\usepackage{hyperref} 
\hypersetup{colorlinks,allcolors=blue!50!black}
\allowdisplaybreaks

\begin{document}

\title{Generalised Marcus Theory for Multi-Molecular Delocalised Charge Transfer}

\author{Natasha B. Taylor}
\affiliation{Centre for Engineered Quantum Systems and School of Mathematics and Physics, The University of Queensland, Queensland 4072, Australia}

\author{Ivan Kassal}
\email{ivan.kassal@sydney.edu.au}
\affiliation{Centre for Engineered Quantum Systems and School of Mathematics and Physics, The University of Queensland, Queensland 4072, Australia}
\affiliation{The University of Sydney Nano Institute and School of Chemistry, The University of Sydney, NSW 2006, Australia}
	
		
\begin{abstract}
	Although Marcus theory is widely used to describe charge transfer in molecular systems, in its usual form it is restricted to transfer from one molecule to another. If a charge is delocalised across multiple donor molecules, this approach requires us to treat the entire donor aggregate as a unified supermolecule, leading to potentially expensive quantum-chemical calculations and making it more difficult to understand how the aggregate components contribute to the overall transfer. Here, we show that it is possible to describe charge transfer between groups of molecules in terms of the properties of the constituent molecules and couplings between them, obviating the need for expensive supermolecular calculations. We use the resulting theory to show that charge delocalisation between molecules in either the donor or acceptor aggregates can enhance the rate of charge transfer through a process we call supertransfer (or suppress it through subtransfer). The rate can also be enhanced above what is possible with a single molecule by judiciously tuning energy levels and reorganisation energies. We also describe bridge-mediated charge transfer between delocalised molecular aggregates. The equations of generalised Marcus theory are in closed form, providing qualitative insight into the impact of delocalisation on charge dynamics in molecular systems.
\end{abstract}
\maketitle

	\section{Introduction}
	Theories of charge-transfer rates underpin our understanding of a wide variety of chemical reactions and charge-transport processes, not only in chemistry, but also in biology and materials science~\cite{Marcus1985a,Barbara1996a,MayKuhn,jortner1999electron}. In most of the well-studied cases, the charge is being transferred from one molecule to another. However, in many systems---including organic semiconductors~\cite{Few2014,kohler2015electronic}, the reaction centres of photosynthetic organisms~\cite{Allen1998,Blankenship}, inorganic coordination complexes~\cite{DAlessandro2006a}, and conductive metal-organic frameworks (MOFs)~\cite{DAlessandro2011}---the charge to be transferred is delocalised across multiple donor molecules (or is to be received by states delocalised over multiple acceptor molecules). The usual theoretical approaches can be applied to these cases if the donor or acceptor aggregates are treated as single supermolecules, but doing so is often computationally prohibitive, requires a complete re-calculation if any part is changed, and, most importantly, offers limited qualitative insight into how the component molecules and the interactions between them affect the inter-aggregate charge transfer.
	
	Although delocalisation in charge transfer has been studied extensively, most studies have focused on cases of delocalisation between the donor and acceptor, as opposed to delocalisation within donor or acceptor aggregates. In particular, donor-acceptor delocalisation is critical to understanding adiabatic electron transfer, as first emphasised by Hush~\cite{Hush1958,Hush1960}, and extended by numerous authors since~\cite{Cave:1996ep,Matyushov2000d,Subotnik:2008fp}. For example, intervalence transitions in mixed-valence compounds are a clear manifestation of delocalisation between two molecules~\cite{hush1967intervalence}. 
	
	Here, we study the problem of charge transfer from one delocalised molecular aggregate to another. In order to be able to speak of two distinct aggregates, we assume that the coupling between the aggregates (i.e., between any donor molecule and any acceptor molecule) is small compared to the strength of their coupling to the environment. Furthermore, to ensure that charges within either aggregate (or both) are delocalised among the constituent molecules, we assume that the couplings between the molecules are stronger than their coupling to the environment. 
	
	Because the overall donor-acceptor coupling is weak, the charge transfer will be incoherent, i.e., with no coherence between the donor and acceptor states. Apart from the delocalisation within the aggregates, this situation is described by non-adiabatic electron transfer, which we take as our starting point. Although we will follow convention in calling it Marcus theory~\cite{Marcus1956a} (MT), the standard expression for non-adiabatic charge transfer between one donor $D$ and one acceptor $A$ was derived by Levich and Dogonadze~\cite{levich1959theory}:
	\begin{equation}\label{ekmarcus}
	k_{D\rightarrow A}=\frac{2\pi}{\hbar}\frac{|V_{DA}|^2}{\sqrt{4\pi k_B T\lambda_{DA}}}\exp\bigg(\frac{-(\Delta E_{DA}+\lambda_{DA})^2}{4k_BT\lambda_{DA}}\bigg),
	\end{equation}
	where, at temperature $T$, three parameters control the transfer rate: the donor-acceptor electronic coupling, $V_{DA}$, determined by the overlap of their electronic wavefunctions; the reorganisation energy, $\lambda_{DA}$, being the energy needed to reorganise the environment from equilibrium about the reactant to equilibrium about the product at fixed electronic state; and the energy difference between the final and initial states, $\Delta E_{DA}$.

	\begin{figure}[t]
		\includegraphics[width=0.95\columnwidth]{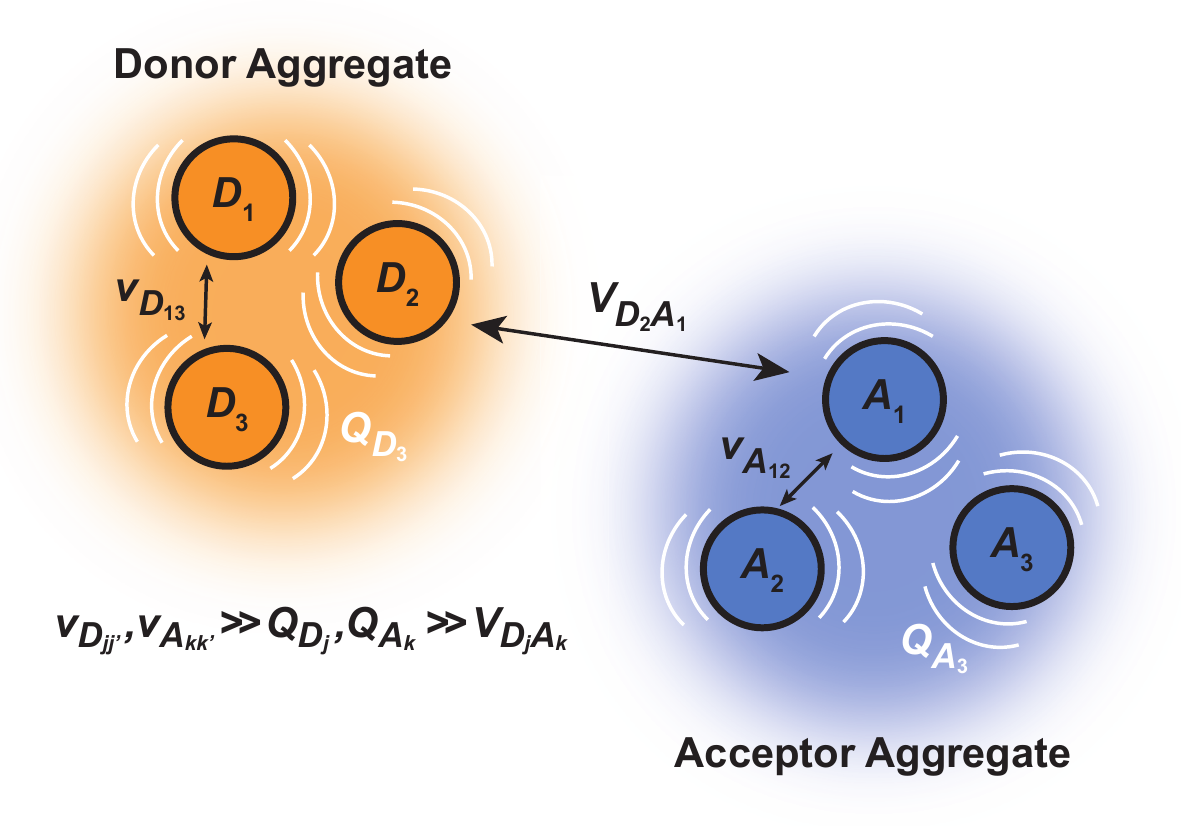}
		\caption{\label{fgmt} The model system for generalised Marcus theory. The model describes charge transfer between two delocalised aggregates if the couplings $v$ between molecules constituting the donor (or acceptor) are strong compared to the coupling to the environment $Q$, while the couplings $V$ between molecules in the donor with those in the acceptor are relatively weak. 
		}
	\end{figure}

	Here, we show that is possible to generalise non-adiabatic MT to describe charge transfer between molecular aggregates in terms of the properties of individual molecules and couplings between them. Our theory is both computationally cheap -- avoiding the need for supramolecular quantum-chemical simulations -- and offers intuitive insight into how the charge transfer rates are affected by changes to molecules in either aggregate. 
	
	Our approach is inspired by developments in Förster resonance energy transfer (FRET), which describes the exciton transfer rate between two chromophores and is, like MT, derived from second-order perturbation theory in the donor-acceptor coupling. Sumi developed generalised FRET (gFRET) to describe the transfer of excitons between delocalised aggregates in photosynthetic antenna complexes~\cite{Sumi1999,Sumi2001a}, and his approach has since been used to study exciton transfer in a wide range of molecular aggregates~\cite{Scholes2001,Baghbanzadeh2016,Baghbanzadeh2016a}. Following Sumi, we name our theory `generalised Marcus theory' (gMT).
	
	MT also allows a description of bridge-mediated charge transfer, where the donor and acceptor are not directly coupled, but a coupling between them is mediated by intervening `bridge' molecules, whose states are sufficiently high in energy to prevent actual charge transfer from the donor to the bridge~\cite{Marcus1985a,Barbara1996a,MayKuhn}. A bridge enables charge transfer to occur over longer distances, although the rate typically decreases exponentially with the number of bridge elements. After deriving gMT in section~\ref{secgmt}, we show that it is also easily extended to describe bridge-assisted charge transfer between delocalised aggregates in section~\ref{sec:bridge}.

	\renewcommand{\arraystretch}{1.5}
\begin{table*}[t]
	\centering
	\begin{tabular}{p{2.5cm}lll} \toprule
		& Marcus theory & Generalised Marcus theory & Bridge-mediated generalised Marcus theory\\ 
		\midrule
		Donor \& acceptor indices & Sites $\ket{D_j}$, $\ket{A_k}$ & Eigenstates $\begin{aligned}[t] \ket{D_\alpha} &= \sum_jc_{\alpha j}\ket{D_j},\\ \ket{A_\beta} &= \sum_kc_{\beta k}\ket{A_k} \end{aligned}$ & Eigenstates $\begin{aligned}[t] \ket{D_\alpha} &= \sum_jc_{\alpha j}\ket{D_j},\\\ket{A_\beta} &= \sum_kc_{\beta k}\ket{A_k} \end{aligned}$\\
		Electronic coupling & $V_{jk}$ & $\displaystyle V_{\alpha\beta}=\sum_{j,k}c_{\alpha j}c_{\beta k}^*V_{jk}$ & $\displaystyle \hat{V}_{\alpha\beta}=\sum_{j,k}c_{\alpha j}c_{\beta k}^* V_{jB_1} G_B^{1N} V_{kB_{N}}$ \\
		Reorganisation energy\newline (separable environments) & $\lambda_{j}+\lambda_{k}$ & $\displaystyle \lambda_{\alpha\beta}=\sum_j |c_{\alpha j}|^4\lambda_{j}+\sum_k |c_{\beta k}|^4\lambda_{k}$ & $\displaystyle \lambda_{\alpha\beta}=\sum_j |c_{\alpha j}|^4\lambda_{j}+\sum_k |c_{\beta k}|^4\lambda_{k}$ \\
		Energy difference $(\Delta E)$ & $E_{A_k}-E_{A_j}$ & $E_{\beta}-E_{\alpha}$ & $\begin{aligned}[t] \hat{E}_{\beta}-\hat{E}_{\alpha} = {} & E_{\beta}+\sum_{k,k^\prime}c_{\beta k}c^\star_{\beta k^\prime}V_{k B_{N}}V_{B_{N} k^\prime}G_ B^{NN} - {} \\ & E_{\alpha}-\sum_{j,j^\prime}c_{\alpha j}c^\star_{\alpha j^\prime}V_{j B_1}V_{B_1 j^\prime}G_ B^{11} \end{aligned}$ \\
		\bottomrule
	\end{tabular}
	\caption{\label{parameters} The equations of generalised Marcus theory (gMT) and bridge-mediated gMT have the same form as ordinary Marcus theory, provided that relevant parameters are replaced as provided in this table. $E_{\alpha}$ and $E_{\beta}$ are the eigenstates of the donor and acceptor aggregate Hamiltonians, $H_D^0$ and $H_A^0$ respectively. Bridge-mediated gMT contains a bridge of $N$ sites, with site $B_1$ coupling to the donor aggregate and $B_{N}$  to the acceptor. The couplings $V_{jB_1}$ and $V_{kB_{N}}$ are the coupling of the $j$th donor site to $B_1$, and the coupling of the $k$th acceptor site to $B_{N}$, respectively. Green's function $G_ B$, given by Equation~\ref{bgreens}, describes transport through the bridge.
	}  
\end{table*}

	\section{Results}
	
	\subsection{Generalised Marcus theory}\label{secgmt}
	
	We generalise Marcus theory by considering an aggregate of $N_D$ donor molecules and an aggregate of $N_A$ acceptor molecules, with each molecule coupled to an independent environment of thermalised harmonic oscillators. Three approximations make it possible to define two distinct aggregates (Fig.~\ref{fgmt}): first, the coupling between molecules in each aggregate is much stronger than their coupling to the environment, ensuring that the delocalised eigenstates of each aggregate are the appropriate basis for perturbation theory; second, the system-environment coupling is much stronger than the inter-aggregate coupling, implying that inter-aggregate charge transfer is incoherent (hopping); and third, because we assume each site is coupled to its own environment, no environmental mode connects a donor and an acceptor molecule. Where applicable, we follow the derivation of multi-chromophoric FRET (MC-FRET)~\cite{Jang2004}, which reduces to generalised FRET in the appropriate limit. While gFRET can also be derived using Fermi's golden rule\cite{Ma2015}, we used a time-dependent derivation because some of our intermediate results may be useful in more general contexts.
	
	The full Hamiltonian is $H=H_D^0+H_A^0+H_C+H_{DE}+H_{AE}+H_E$, and we introduce each term here as well as in Fig.~\ref{fgmt}. The donor-aggregate and acceptor-aggregate Hamiltonians are, respectively,
	\begin{align}
	H_D^0&=\sum_{j=1}^{N_D}E_{j}\ket{D_j}\bra{D_j}+\sum_{j\neq j'}v_{jj'}\ket{D_j}\bra{D_{j'}},\\
	H_A^0&=\sum_{k=1}^{N_A}E_{k}\ket{A_k}\bra{A_k}+\sum_{k\neq k'}v_{kk'}\ket{A_k}\bra{A_{k'}},
	\end{align}
	where $\ket{D_j}$ and $\ket{A_k}$ are the states where the charge is localised on molecules $D_j$ and $A_k$ respectively. Throughout this work we index donor sites with $j$ and acceptor sites with $k$: $\sum_j$ should be read as a sum over only the donor sites, and $\sum_k$ only over acceptors. The donor and acceptor molecules have site energies $E_{j}$ and $E_{k}$, and intra-aggregate couplings are $v_{jj'}$ (in the donor) and $v_{kk'}$ (in the acceptor). 
	
	We refer to the eigenstates of $H_D^0$ and $H_A^0$ as the aggregate basis, being, respectively, $\ket{D_\alpha}=\sum_j c_{\alpha j}\ket{D_j}$ and $\ket{A_\beta}=\sum_k c_{\beta k}\ket{A_k}$, with energies $E_{\alpha}$ and $ E_{\beta}$. Similar to site indices $j$ and $k$, index $\alpha$ is consistently used to denote only donor eigenstates, and $\beta$ acceptor eigenstates.
	
	Inter-aggregate coupling is described by the Hamiltonian 
	\begin{equation} 
	H_C=\sum_{j=1}^{N_D}\sum_{k=1}^{N_A}V_{jk}\big(\ket{D_j}\bra{A_k}+\ket{A_k}\bra{D_j}\big),
	\end{equation}
	where $V_{jk}$ is the coupling between the $j$th donor and $k$th acceptor molecules.	
	
	The environment is described by a set of harmonic oscillators:
	\begin{equation}
	H_{E}=\sum_\xi\hbar\omega_{\xi}(b_{\xi}^\dagger b_{\xi}+1/2),
	\end{equation}
	where $\omega_{\xi}$ is the frequency of the $\xi$th environment mode, with creation operator $b_{\xi}^\dagger$. We can also write $H_E=H_{E_D} + H_{E_A}$, with the environment modes partitioned between those that couple to donor and acceptor molecules.
	
	The donor-environment and acceptor-environment interaction Hamiltonians are, respectively,
	\begin{align}
	H_{DE}&=\sum_{j=1}^{N_D}Q_{j}\ket{D_j}\bra{D_j},\\
	H_{AE}&=\sum_{k=1}^{N_A}Q_{k}\ket{A_k}\bra{A_k},
	\end{align}
	with $Q_{j}=\sum_{\xi}\hbar\omega_\xi g_{j\xi} (b_{\xi}+b_{\xi}^\dagger)$, where $g_{j\xi}$ is the dimensionless coupling of the $\xi$th environment mode to the charged $j$th donor molecule, relative to the uncharged state. $Q_{k}$ is defined analogously. The assumption of a local environment means that, for a fixed $\xi$, only one of $g_{j\xi}$ can be non-zero.
	
	The charge-transfer rate is the rate of change of the charge population on the acceptor,
	\begin{equation}\label{ebegin}
	k_{D\rightarrow A}(t)=\frac{d}{dt}\mathrm{Tr}_E\sum_{k}\bra{A_k}\rho(t)\ket{A_k},
	\end{equation}
	where $\rho(t)$ is the density matrix of the system, and $\mathrm{Tr}_E$ is the trace over the environmental modes. As detailed in the Appendix, $k_{D\rightarrow A}$ can be calculated using second-order perturbation theory in $H_C$ and, because we assumed separable environments, generates a time-dependent transfer rate
	\begin{multline}\label{ieexpandedform}
	k_{D\rightarrow A}(t)=\sum_{j,j'}\sum_{k,k'}\frac{V_{jk}V_{j' k'}}{\hbar^2}\cdot 2\mathrm{Re}
	\int_0^t d\tau\,\mathrm{Tr}_E\big(\\
	\Bra{A_k}e^{-i(H-H_C)(t-\tau)/\hbar}\Ket{A_{k'}}\Bra{D_{j'}}e^{-i(H-H_C)\tau/\hbar}\\\rho(0)e^{i(H-H_C)t/\hbar}\Ket{D_j}
	\big).
	\end{multline}
	
	To proceed, we consider the rate in the aggregate basis. The requirement that $V_{jk}$ be weaker than all other couplings means that the donor aggregate will relax to a thermal state faster than the charge transfer. In other words, we assume that the initial density operator of the system $\rho(0)$ will, before charge transfer takes place, relax to a state $\rho_\mathrm{th}$ in which both the donor and acceptor aggregates are in equilibrium with their own environments (see Appendix for details). This gives a time-independent transfer rate,
	\begin{equation}\label{iediag}
	k_{D\rightarrow A}=\sum_{\alpha,\beta}\frac{|V_{\alpha\beta}|^2}{2\pi\hbar^2}\int_{-\infty}^\infty d\omega\,\mathcal{D}_D^{\alpha\alpha}(\omega)\mathcal{A}_A^{\beta\beta}(\omega),
	\end{equation}
	where
	\begin{align}
	V_{\alpha\beta}=&\sum_{j,k}c^*_{\alpha j}c_{\beta k}V_{jk},\\
	\label{ieDfinal}
	\mathcal{D}_D^{\alpha\alpha}(\omega)=&\int_{-\infty}^{\infty}dt\,e^{-i\omega t}\mathrm{Tr}_{E_D}\big(\nonumber\\&e^{-iH_{E_D}t/\hbar}
	\Bra{D_{\alpha}}e^{iH_D t/\hbar}\rho_D\Ket{D_{\alpha}}\big),\\
	\label{ieAfinal}
	\mathcal{A}_A^{\beta\beta}(\omega)=&\int_{-\infty}^{\infty}dt\,e^{i\omega t}\mathrm{Tr}_{E_A}\big(\nonumber\\&e^{iH_{E_A}t/\hbar}
	\Bra{A_{\beta}}e^{-iH_A t/\hbar}\Ket{A_{\beta}}\rho_A\big),
	\end{align}
	and where $\rho_\mathrm{th}$ is split into donor and acceptor components, $\rho_\mathrm{th}=\rho_D\otimes\rho_A$. Because the donor-environment coupling is weak, the thermal state of the donor will approximately factorise to $\rho_D=(\sum_{\alpha}\rho_{\alpha\alpha}\ket{D_\alpha}\bra{D_\alpha})\otimes\rho_{E_D}$, where the electronic population distribution is $\rho_{\alpha\alpha}=\exp(-E_\alpha/k_BT)/(\sum_{\alpha}\exp(-E_\alpha/k_BT))$ and the thermal environment is $\rho_{E_D}=\exp(-H_{E_D}/k_BT)/\mathrm{Tr}_{E_D}(\exp(-H_{E_D}/k_BT))$. The thermal state of the acceptor is $\rho_A=\rho_{E_A}=\exp(-H_{E_A}/k_BT)/\mathrm{Tr}_{E_A}(\exp(-H_{E_A}/k_BT))$. Finally, we have also written $H_D=H_D^0+H_{DE}+H_{E_D}$, and similarly for $H_A$.
	
	Eqs.~\ref{iediag}--\ref{ieAfinal} are analogous to the MC-FRET treatment of delocalised exciton transfer~\cite{Jang2004}. In particular, the rate of MC-FRET depends on the (weighted) overlap of the donor emission spectrum with the acceptor absorption spectrum, which resembles the form of Eq.~\ref{iediag}. However, in gMT, Eq.~\ref{ieDfinal} describes the spectrum of charge disassociation from the donor and Eq.~\ref{ieAfinal} the charge association spectrum for the acceptor. Furthermore, the inter-aggregate coupling in Eq.~\ref{iediag} is determined by the overlap of electronic wavefunctions, while in MC-FRET the couplings are from the interactions of transition dipole moments. 
	
	Evaluating Eqs.~\ref{ieDfinal}--\ref{ieAfinal} for independent harmonic environments gives (see Appendix for details) 
	\begin{align}
	\label{iedint}\mathcal{D}_D^{\alpha\alpha}(\omega)&=\rho_{\alpha\alpha}\int_{-\infty}^{\infty}dt\,e^{-i\omega t}e^{iE_\alpha t/\hbar+G_{\alpha}(t)-G_{\alpha}(0)},\\
	\label{ieaint}\mathcal{A}_A^{\beta\beta}(\omega)&=\int_{-\infty}^{\infty}dt\,e^{i\omega t}e^{-iE_\beta t/\hbar+G_\beta(t)-G_\beta(0)},
	\end{align}
	with the lineshape function 
	\begin{equation}\label{Gtrigform}
	G_{\alpha}(t)=\sum_{\xi}g_{\alpha\xi}^2\big(\cos(\omega_{\xi}t)(1+2n(\omega_{\xi}))-i\sin(\omega_{\xi}t)\big),
	\end{equation}
	and $G_{\beta}(t)$ analogously defined. For a thermally populated environment, the occupation of environmental modes is given by the Bose-Einstein distribution $n(\nu)=(\exp(\hbar\nu/k_BT)-1)^{-1}$.
	
	The preceding equations are appropriate at a wide range of temperatures and environmental spectral densities. Although we could stop here, to obtain a clear comparison with MT, we now make two additional approximations that are also made in deriving ordinary Marcus theory. To do so, we assume that the spectral density $J(\omega)=\sum_{\xi}g_{\alpha\xi}^2 \delta(\omega-\omega_\xi)$ goes rapidly to zero beyond a cut-off frequency $\omega_c$. Then, we first assume the high-temperature limit $k_BT\gg\hbar\omega_c$, so that $n(\nu)\approx k_BT/\hbar\nu\gg1$, giving
	\begin{equation}\label{Gtrigform2}
	G_{\alpha}(t)=\sum_{\xi}g_{\alpha\xi}^2\bigg(\frac{2k_BT}{\hbar\omega_{\xi}} \cos(\omega_{\xi}t)-i\sin(\omega_{\xi}t)\bigg),
	\end{equation}
	Second, MT also assumes the slow-nuclear-mode limit, in which the charge-transfer occurs faster than the characteristic timescales of the environment: $t\ll 1/\omega_c\lessapprox 1/\omega_{\xi}$. With $\omega_{\xi}t\ll 1$, we expand the trigonometric functions in Eq.~\ref{Gtrigform2} to leading order:
	\begin{equation}\label{Gtrigform3}
	G_{\alpha}(t)=\sum_{\xi}g_{\alpha\xi}^2\bigg(\frac{2k_BT}{\hbar\omega_{\xi}}-\frac{t^2k_BT\omega_{\xi}}{\hbar} -i\omega_{\xi}t\bigg).
	\end{equation}
	
	We now define the reorganisation energy for the donor sites as $\lambda_j=\sum_{\xi}\hbar\omega_{\xi}g_{j\xi}^2$, and similarly for the acceptor sites, $\lambda_k$. The change of basis $g_{\alpha\xi}=\sum_j|c_{\alpha j}|^2 g_{j\xi}$ gives the reorganisation energy of aggregate eigenstates
	\begin{align}
	\lambda_{\alpha} &= \sum_{\xi}\hbar\omega_{\xi}g_{\alpha\xi}^2 \nonumber
	\\&= \sum_{\xi}\hbar\omega_{\xi}\Big(\sum_j|c_{\alpha j}|^2 g_{j\xi}\Big)\Big(\sum_{j'}|c_{\alpha j'}|^2 g_{j'\xi}\Big). \label{reorgcalc}
	\end{align} 
	Since each site has an independent environment, no mode $\xi$ couples to two different sites ($g_{j\xi}g_{j'\xi}=g_{j\xi}^2\delta_{jj'}$), giving
	\begin{equation}\label{reorgfinal}
	\lambda_{\alpha} = \sum_{\xi}\hbar\omega_{\xi}\sum_j|c_{\alpha j}|^4 g_{j\xi}^2,
	\end{equation}
	and similarly for $\lambda_\beta$.
	
	Substituting Eqs.~\ref{Gtrigform3} and \ref{reorgfinal} into Eqs.~\ref{iedint}--\ref{ieaint} we find
	\begin{align}
	\mathcal{D}_D^{\alpha\alpha}(\omega)&= \rho_{\alpha\alpha}\frac{2\pi\hbar}{\sqrt{4\pi k_BT\lambda_{\alpha}}}\exp\bigg(\frac{-(E_{\alpha}-\hbar\omega-\lambda_{\alpha})^2}{4k_BT\lambda_{\alpha}}\bigg),\\
	\mathcal{A}_A^{\beta\beta}(\omega)&=\frac{2\pi\hbar}{\sqrt{4\pi k_BT\lambda_{\beta}}}\exp\bigg(\frac{-(E_{\beta}-\hbar\omega+\lambda_{\beta})^2}{4k_BT\lambda_{\beta}}\bigg).
	\end{align}
	Consequently, the overlap integral in Eq.~\ref{iediag} becomes
	\begin{equation}\label{result}
	k_{D\rightarrow A}=\sum_{\alpha,\beta}\frac{2\pi}{\hbar}\frac{\rho_{\alpha\alpha}|V_{\alpha\beta}|^2}{\sqrt{4\pi k_B T\lambda_{\alpha \beta}}}
	\exp\bigg(\frac{-(\Delta E_{\alpha \beta}+\lambda_{\alpha \beta})^2}{4k_BT\lambda_{\alpha \beta}}\bigg),
	\end{equation}
	where $\Delta E_{\alpha\beta}=E_\beta-E_\alpha$ and $\lambda_{\alpha\beta}=\lambda_\alpha+\lambda_\beta$, demonstrating that gMT takes the same form as MT, with all parameters defined analogously to---and expressible in terms of---their site-basis counterparts. These results are also summarised in Table~\ref{parameters}, and in the limit of a single-molecule donor and single-molecule acceptor, Eq.~\ref{result} reduces to the ordinary MT rate, Eq.~\ref{ekmarcus}. The ability to recast gFRET in a form analogous to Equation~\ref{result}~\cite{Cleary2013} further illustrates the deep similarities between charge and exciton transfer.

	\begin{figure}[!t]
		\includegraphics[width=0.95\columnwidth]{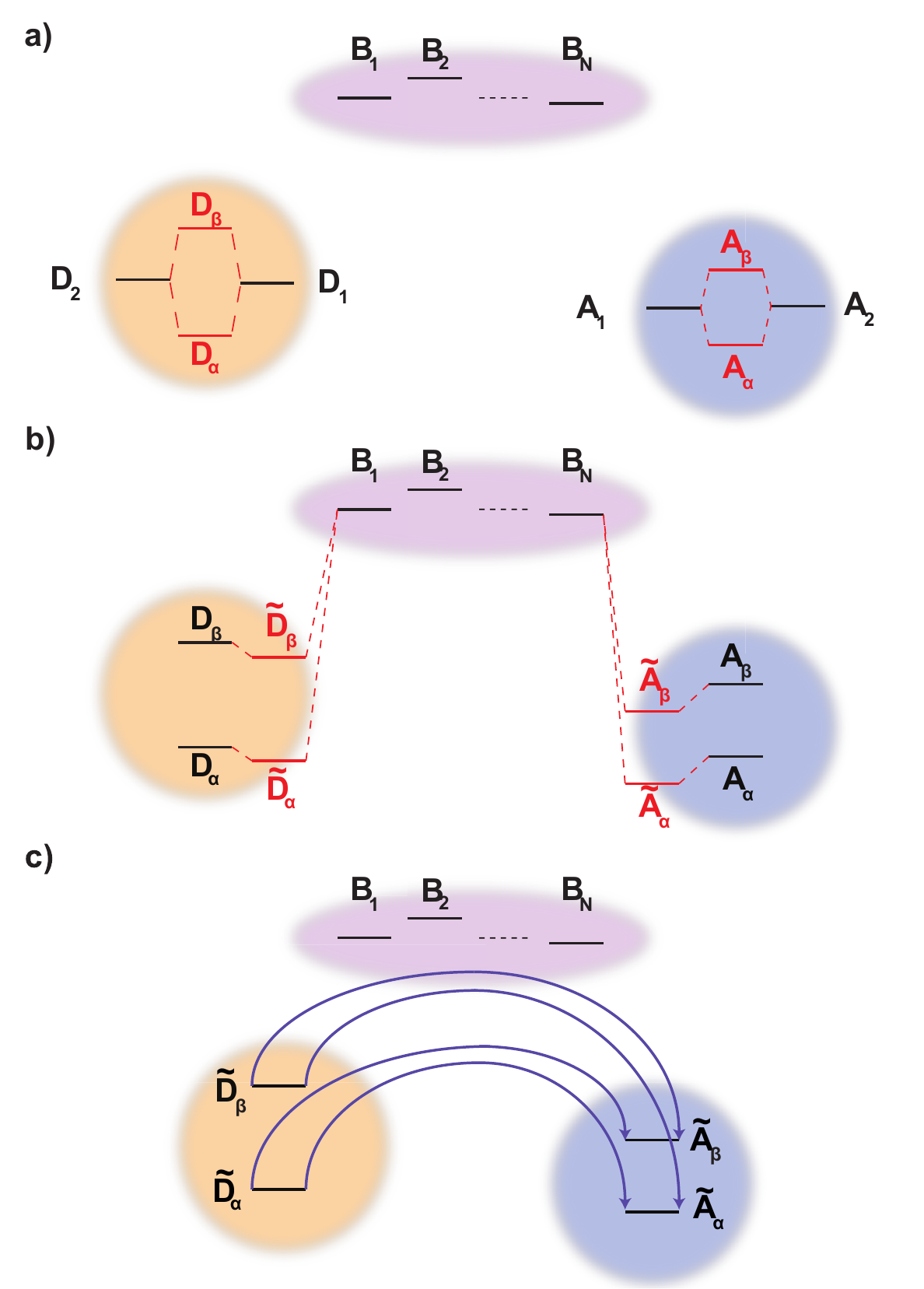}
		\caption{\label{fbridge} Generalised bridge-assisted charge transfer, shown with two donor molecules, $D_1$ and $D_2$, two acceptor molecules, $A_1$ and $A_2$, and $N$ bridge molecules, $B_1$ \ldots $B_N$. a) The eigenstates of each aggregate are calculated. b) The energies of these eigenstates are then perturbed by the coupling to the bridge (perturbation of the bridge levels is neglected, see text). c) Charge transfer occurs directly between donor and acceptor aggregate eigenstates, assisted by virtual bridge states.		
		}
	\end{figure}

	\subsection{Generalised bridge-mediated charge transfer}\label{sec:bridge}
	
	Like MT, gMT can be expanded to include the case where the coupling between the donor and the acceptor aggregates is not direct, but is instead mediated by a bridge consisting of higher-lying states of intervening molecules. We consider a bridge of $N$ molecules, each modelled as a single site, where the donor molecules only couple to the first bridge state, $B_1$, the acceptor molecules only couple to the last bridge state, $B_N$, and each bridge molecule only couples to its two nearest neighbours in the chain, as shown in Fig.~\ref{fbridge}. Usually, the bridge site energies $E_{B_l}$ are considered energetically distinct from the donor and acceptor aggregates, i.e.,
	\begin{equation}\label{inequal}
	(E_{B_k}-E_{B_l}),|V_{B_l B_{l+1}}|\ll E_{B_l}-E_{\alpha/\beta},
	\end{equation}	
	where $V_{B_l B_{l+1}}$ are the intra-bridge couplings and $E_{\alpha/\beta}$ is the characteristic energy of donor and acceptor eigenstates (for concreteness, it could be taken as the highest eigenvalue of either $H_D^0$ or $H_A^0$).
	
	We define the donor-bridge-acceptor Hamiltonian $H_{DBA}=H_B + H_D^0 + H_{DB}+ H_A^0 + H_{AB}$ using $H_D^0$ and $H_A^0$ as above and adding the bridge Hamiltonian $H_B$ and the coupling of the bridge to the donor, $H_{DB}$, and acceptor, $H_{AB}$,
	\begin{align}
	H_B=&\sum_{l=1}^N E_{B_l}\Ket{B_l}\Bra{B_l}\nonumber\\&+\sum_{l=1}^{N-1} V_{B_l B_{l+1}} \Ket{B_l}\Bra{B_{l+1}}+\mathrm{h.c.},\\
	H_{DB}=&\sum_j V_{j B_1} \Ket{D_j}\Bra{B_1}+\mathrm{h.c.},\\
	H_{AB}=&\sum_k V_{k B_N}\Ket{A_k}\Bra{B_N}+\mathrm{h.c.},
	\end{align}
	where $\ket{B_l}$ is the state of a charge being located on bridge site $B_l$.
	
	Instead of thinking of $B_1$ and $B_N$ as coupling to donor and acceptor sites, we can also consider them as coupling to the aggregate eigenstates. In the aggregate basis, $H_{DBA}$ becomes
	\begin{multline}
	H_{DBA}= \sum_{\alpha}\big(E_{\alpha}\Ket{D_\alpha}\Bra{D_\alpha}+V_{\alpha B_1}\Ket{D_\alpha}\Bra{B_1}+\mathrm{h.c.}\big)\\
	+\sum_\beta \big(E_{\beta}\Ket{A_\beta}\Bra{A_\beta}+V_{\beta B_N}\Ket{A_\beta}\Bra{B_N}+\mathrm{h.c.}\big)+H_{B},
	\end{multline}
	where $V_{\alpha B_1}=\sum_j c_{\alpha j}V_{j B_1}$ and $V_{\beta B_N}=\sum_k c_{\beta k}V_{k B_N}$.
	
	We calculate the rate of charge transfer from each donor eigenstate $\ket{D_\alpha}$ to each acceptor eigenstate $\ket{A_\beta}$ independently, using the mathematics already established for single-site bridge-mediated transfer~\cite{Nitzan}. In other words, instead of considering the entire donor-bridge-acceptor system, we consider separately the subspace of each donor and acceptor eigenstate with the bridge, 
	\begin{align}
	H_{DBA}(\alpha,\beta)={}& E_{\alpha}\Ket{D_\alpha}\Bra{D_\alpha}+V_{\alpha B_1}\Ket{D_\alpha}\Bra{B_1}+\mathrm{h.c.}\nonumber\\
	&+E_{\beta}\Ket{A_\beta}\Bra{A_\beta}+V_{\beta B_N}\Ket{A_\beta}\Bra{B_N}+\mathrm{h.c.}\nonumber\\&+H_{B}
	\end{align}
	We denote the lowest-eigenvalue eigenvector of $H_{DBA}(\alpha,\beta)$ as $\mathbf{d}_{DBA}=(d_{\alpha},d_{B_1},\ldots,d_{B_N},d_{\beta})$, with eigenvalue $E_{DBA}$. 
	
	Since  $(H_{DBA}(\alpha,\beta)-\mathbf{I}E_{DBA})\mathbf{d}_{DBA}=0$, we find that
	\begin{align}\label{inhomo1}
	(E_{\alpha}-E_{DBA})d_{\alpha}+V_{\alpha B_1} d_{B_1}&=0,\\\label{inhomo2}
	(E_{\beta}-E_{DBA})d_{\beta}+V_{\beta B_N} d_{B_N}&=0.
	\end{align}
	The values of $d_{B_1}$ and $d_{B_N}$ can be found by considering the bridge subspace, $(H_B-\mathbf{I}E_{DBA}) \mathbf{d}_B = -(V_{B_1\alpha}d_{\alpha},0,\ldots,0,V_{B_N\beta}d_{\beta})$ where $\mathbf{d}_B$ consists of the bridge elements of $\mathbf{d}_{DBA}$ in the same order. The solution of this equation is $\mathbf{d}_B= G_B(V_{B_1\alpha}d_{\alpha},0,\ldots,0,V_{B_N\beta}d_{\beta})$, using Green's function $G_B=(\mathbf{I}E_{DBA}-H_B)^{-1}$.
	
	By substituting this solution for $d_{B_1}$ and $d_{B_N}$ into Equations \ref{inhomo1} and \ref{inhomo2}, we find
	\begin{align}
	(\hat{E}_\alpha-E_{DBA})d_{\alpha}+\hat{V}_{\alpha \beta} d_{\beta}&=0,\\
	(\hat{E}_\beta-E_{DBA})d_{\beta}+\hat{V}_{\beta \alpha} d_{\alpha}&=0,
	\end{align}
	where $\hat{E}$ are the perturbed energies of aggregate eigenstates due to coupling with the bridge,
	\begin{align}\label{btransforms}
	\hat{E}_{\alpha} &= E_{\alpha}+V_{\alpha B_1}G_ B^{11}V_{B_1 \alpha}, \\
	\hat{E}_{\beta} &= E_{\beta}+V_{\beta B_N}G_ B^{NN}V_{B_N\beta},
	\end{align}
	and $\hat{V}$ is the effective coupling between the donor and acceptor eigenstates, mediated by the bridge,
	\begin{equation}\label{bcouplings}
	\hat{V}_{\alpha \beta} = V_{\alpha B_1}G_ B^{1N}V_{B_N\beta}.
	\end{equation}
	
	To find the Green's function, we expand $G_B$ in terms of a Dyson series. Because $|V_{B_l B_{l+1}}|$ is small (see Eq.~\ref{inequal}), we keep only the lowest-order term~\cite{Nitzan},
	\begin{multline}\label{bgreens}
	G_ B^{1N}=(E_{DBA}-E_{B_1})^{-1}V_{B_1 B_2}(E_{DBA}-E_{B_2})^{-1}\times\\V_{B_2 B_3} \cdots V_{B_{N-1} B_N}(E_{DBA}-E_{B_N})^{-1}.
	\end{multline}
	While $E_{DBA}$ is an eigenvalue of the entire donor-bridge-acceptor system, we are only interested in the donor/acceptor subspace. Because $E_{B_l}-E_{\alpha/\beta}$ is large relative to inter-site couplings and energy differences (Eq.~\ref{inequal}), we can approximate $E_{DBA}-E_{B_l}\approx E_{\alpha/\beta}-E_B$, for average bridge energy $E_B$. This allows us to simplify Eq.~\ref{bcouplings} using the geometric mean of the bridge couplings $V_{BB}$,
	\begin{equation}\label{BV}
	\hat{V}_{\alpha \beta} = \frac{V_{\alpha B_1}V_{B_N\beta}}{E_{\alpha/\beta}-E_B}\bigg(\frac{V_{BB}}{E_{\alpha/\beta}-E_B}\bigg)^{N-1}.
	\end{equation}
	As in ordinary bridge-assisted charge transfer, the effective coupling decays exponentially with bridge length because $V_{BB}<E_{\alpha/\beta}-E_B$ (Eq.~\ref{inequal}). Substituting Eqs.~\ref{btransforms}--\ref{bgreens} into Eq.~\ref{result}, we have the rate of bridge-assisted gMT:
	\begin{equation}\label{bridgeresult}
	k_{D\rightarrow A}=\sum_{\alpha,\beta}\frac{2\pi}{\hbar}\frac{\rho_{\alpha\alpha}|\hat{V}_{\alpha\beta}|^2}{\sqrt{4\pi k_B T\lambda_{\alpha \beta}}}\exp\bigg(\frac{-(\Delta \hat{E}_{\alpha \beta}+\lambda_{\alpha \beta})^2}{4k_BT\lambda_{\alpha \beta}}\bigg).
	\end{equation}

	\section{Discussion}\label{discussion}
	
	The summary of results in Table~\ref{parameters} shows that gMT---whether bridged or not---follows the same functional form as ordinary Marcus theory. This allows intuition gained from studying MT to continue to be useful when studying aggregates instead of single molecules (provided that the parameters are redefined as shown in Table~\ref{parameters}). Further, gMT allows known values of relevant parameters (couplings, energy differences, and reorganisation energies) of individual molecules to be used to calculate the effective parameters for aggregates, saving computational time by avoiding expensive supramolecular quantum-chemical simulations. 
	
	However, the presence of delocalisation in aggregates leads to significant differences between MT and gMT. We can analyse the influence of delocalisation on charge transfer by separating its impact on the electronic and nuclear components of the MT rate.
	
	The gMT electronic coupling factor $|V_{\alpha\beta}|^2=|\sum_{j,k}c_{\alpha j}c_{\beta k}^*V_{jk}|^2$ includes a coherent sum involving electronic amplitudes in each of the donor and acceptor aggregates, allowing both constructive and destructive interference to affect the transfer rate. If the interferences is constructive, leading to enhanced transfer rates, we call the effect \emph{supertransfer}, and if it is destructive, \emph{subtransfer}, borrowing terminology from the similar problem of MC-FRET~\cite{Jang2013a}. 	
	
	For illustration, we consider an aggregate of two identical coupled donors, $D_1$ and $D_2$, with a charge delocalised between them in the $\ket{D_\alpha}=(\ket{D_1}+\ket{D_2})/\sqrt{2}$ state. The donors are coupled to a single acceptor $A$ with strengths $V_{D_1A}$ and $V_{D_2A}$ respectively. If we were to apply Marcus theory between each donor and the acceptor independently, we would expect a transfer rate proportional to the square of each coupling, $k_{MT}\propto \frac{1}{2}|V_{D_1A}|^2+\frac{1}{2}|V_{D_2A}|^2$, with the factors of $1/2$ indicating the population on each donor. However, this naive approach fails to include coherent effects of the superposition. These are treated correctly by gMT, which predicts a transfer rate of $k_{gMT}\propto |(V_{D_1A}+V_{D_2A})/\sqrt{2}|^2$. The presence of rate-enhancement due to supertransfer is apparent if $V_{D_1A}=V_{D_2A}$, which implies $k_{gMT}=2k_{MT}$. In contrast, if the two transfer pathways interfere destructively, $V_{D_1A}=-V_{D_2A}$, gMT predicts subtransfer with $k_{gMT}=0$. We refer to states that enhance the charge-transfer rate through supertransfer as \emph{bright}, while those that retard it as \emph{dark}, in analogy to the terms used in the literature on superradiance~\cite{Fruchtman2016}. The relative populations of the bright and dark states will strongly influence the rate of charge transfer in delocalised systems.

	\begin{figure}[!t]
	\includegraphics{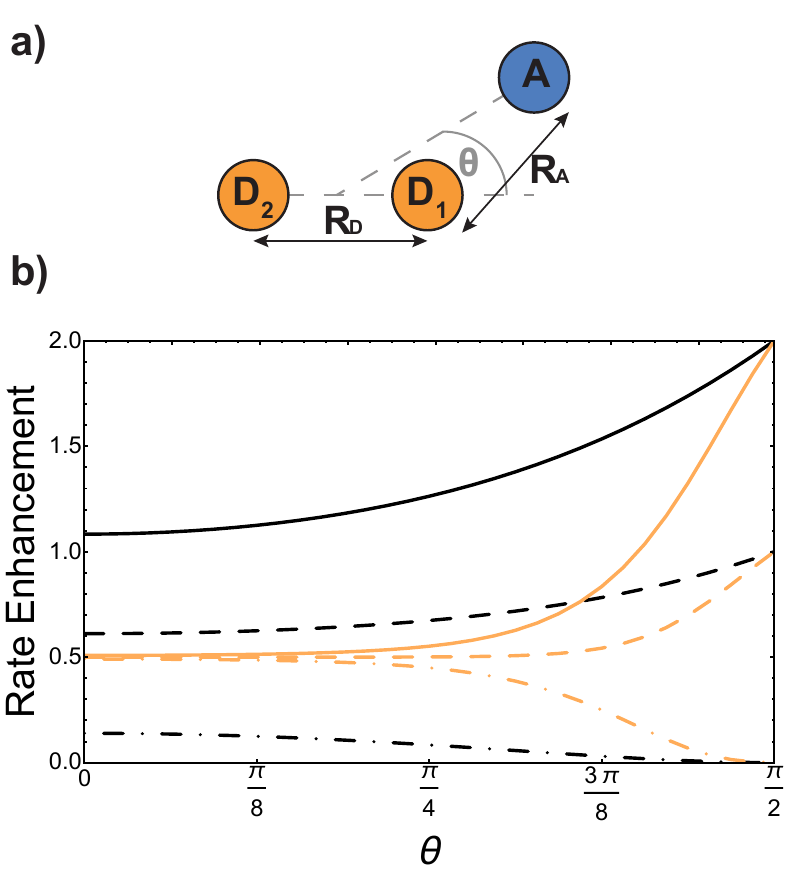}
	\caption{\label{fangle} Example of generalised Marcus theory (gMT) and supertransfer, showing only the impact of electronic component $|V_{\alpha\beta}|^2$ on the charge transfer rate.
		a) Geometric arrangement of two donors and one acceptor, changing from collinear ($\theta=0$) to an isosceles triangle ($\theta=\pi/2$). b) Rates of charge transfer from the donors to the acceptor are displayed as ratios of the rate that would be found if only donor $D_1$ were present and the charge initially localised on it. Black and orange lines indicate, respectively, geometries with $R_A=3R_D$ or $R_A=R_D$ (at a constant $R_A=\SI{5}{\angstrom}$). In both cases, the rates are computed for three initial donor states: the bright state $(\ket{D_1}+\ket{D_2})/\sqrt{2}$ (solid), the dark state $(\ket{D_1}-\ket{D_2})/\sqrt{2}$ (dot-dashed), and the fully mixed state of $\ket{D_1}$ and $\ket{D_2}$ (dashed). These three states are obtained as ground states of the donor Hamiltonian by assuming $V_{D_1D_2}=\SI{-100}{\milli\electronvolt}$ (bright), $V_{D_1D_2}=\SI{100}{\milli\electronvolt}$ (dark), or $V_{D_1D_2}=\SI{0}{\milli\electronvolt}$ (mixed).
		The transfer rates are independent of $R_A$ and $R_D$ when $\theta$ reaches $\pi/2$, where both donors are equidistant from the acceptor. At that point, constructive interference ensures that the transfer from the bright state is twice as fast as it would be from either site alone, while transfer from the dark state is completely suppressed by destructive interference caused by the opposite signs of the wavefunction at $D_1$ and $D_2$. The difference between the two geometries is apparent at smaller $\theta$. When $R_D$ is large compared to $R_A$ (orange), the rate is half the single-site rate for all initial states, indicating that the acceptor is interacting primarily with $D_1$ until $\theta$ becomes considerable. By contrast, when both donors are close enough to the acceptor to interact with it strongly (black), supertransfer and subtransfer can occur at all values of $\theta$, resulting in rate enhancements different from 0.5 at all angles.
		Other calculation parameters: $V_{DA}(r)=\SI{50}{\milli\electronvolt} \exp(1-r/\SI{2}{\angstrom})$. 
	}
\end{figure}

	Supertransfer is also sensitive to the system's geometry. Changing the distance and orientations between donors and acceptors will affect the electronic wavefunction overlaps due to the exponential decay of electronic wavefunctions with distance, consequently modifying the electronic couplings. To explore the consequences of this geometric sensitivity, we consider a model consisting of two donor molecules transferring a unit of charge to an acceptor molecule, shown in Fig.~\ref{fangle}. These calculations demonstrate that rate enhancement/retardation is weakest when the acceptor is co-linear with the donors. This is because the farther donor is so far away that the acceptor is only affected by the nearer donor. The impact is most significant when the acceptor is equidistant from the two donors, where supertransfer from the bright state amplifies the transfer rate by a factor of two, while the dark state provides no transfer.
	
	We can compare these results with gFRET, the analogous theory of excitation-energy transfer between molecular aggregates~\cite{Jang2004}. Bright and dark states also exist in gFRET, but exciton transfer is not as sensitive to small changes in the separation between molecules. While the transfer rate in gMT is determined by the overlap of electronic wavefunctions, which decay exponentially with distance, the MC-FRET rate depends on the coupling of transition dipole moments, which decays with the cube of the distance. In addition, both gMT and gFRET are strongly affected by the relative orientations of the molecules. The orientational dependence of gFRET is easier to predict, especially in the large-separation limit where it can be represented by the interaction of two dipoles. By contrast, the orientational dependence of electronic couplings depends on the shape of the orbitals, which varies from molecule to molecule. Given that the geometric dependence of gFRET can lead to substantially different outcomes in light-harvesting complexes~\cite{Baghbanzadeh2016,Baghbanzadeh2016a}, the stronger dependence of gMT on geometry provides an opportunity to engineer molecular systems that perform charge transfer better than single sites.
	
	The nuclear factor in gMT (also referred to as the Franck-Condon weighted density), $(4\pi k_B T\lambda_{\alpha \beta})^{-1/2}\exp(-(\Delta E_{\alpha \beta}+\lambda_{\alpha \beta})^2/4k_BT\lambda_{\alpha \beta})$, has several features in common with ordinary MT. For example, for a fixed $\lambda_{\alpha\beta}$, the nuclear factor is maximised when $-\Delta E_{\alpha\beta}=\lambda_{\alpha\beta}$, and the inverted regime is possible when $-\Delta E_{\alpha\beta}>\lambda_{\alpha\beta}$. However, the nuclear term also possesses features not predicted by ordinary MT, allowing for both enhancement or retardation of the transfer rate. 
	
	The nuclear factor depends on two energies, $\Delta E_{\alpha \beta}$ and $\lambda_{\alpha \beta}$, which are affected by delocalisation in different ways. On the one hand, $\Delta E_{\alpha \beta}$ is the difference between eigenvalues of $H_D^0$ and $H_A^0$. If the extent of delocalisation in, say, the donor is increased, $E_\alpha$ will not change dramatically, remaining close (up to several times the intermolecular coupling) to a value of typical site energies. On the other hand, $\lambda_{\alpha \beta}$ is reduced by delocalisation. Since $\lambda_\alpha=\sum_j|c_{\alpha j}|^4\lambda_j$, for a state purely localised on $j$, $\lambda_\alpha=\lambda_j$. However, in a fully delocalised state ($c_{\alpha j}=1/\sqrt{N_D}$) of $N_D$ identical donors ($\lambda_j=\lambda$), the reorganisation energy is decreased $N_D$-fold:
	\begin{equation}
	\lambda_\alpha=\sum_{j=1}^{N_D}\bigg|\frac{1}{\sqrt{N_D}}\bigg|^4\lambda_j=\frac{\lambda}{N_D}.
	\end{equation}
	In general, the reduction is by a factor equal to the inverse participation ratio $\mathrm{IPR} = (\sum_j|c_{\alpha j}|^4)^{-1}$. A reduction in $\lambda$ leads to an exponential narrowing of both $\mathcal{D}_D^{\alpha\alpha}(\omega)$ and $\mathcal{A}_A^{\beta\beta}(\omega)$. Therefore, because the charge-transfer rate depends on the overlap of the two spectra (Equation~\ref{iediag}), the reduction in $\lambda$ will reduce the transfer rate between most pairs of eigenstates, the exception being ones where $\Delta E_{\alpha \beta}=-\lambda_{\alpha \beta}$.
	
\begin{figure}[!t]
		\includegraphics[width=0.95\columnwidth]{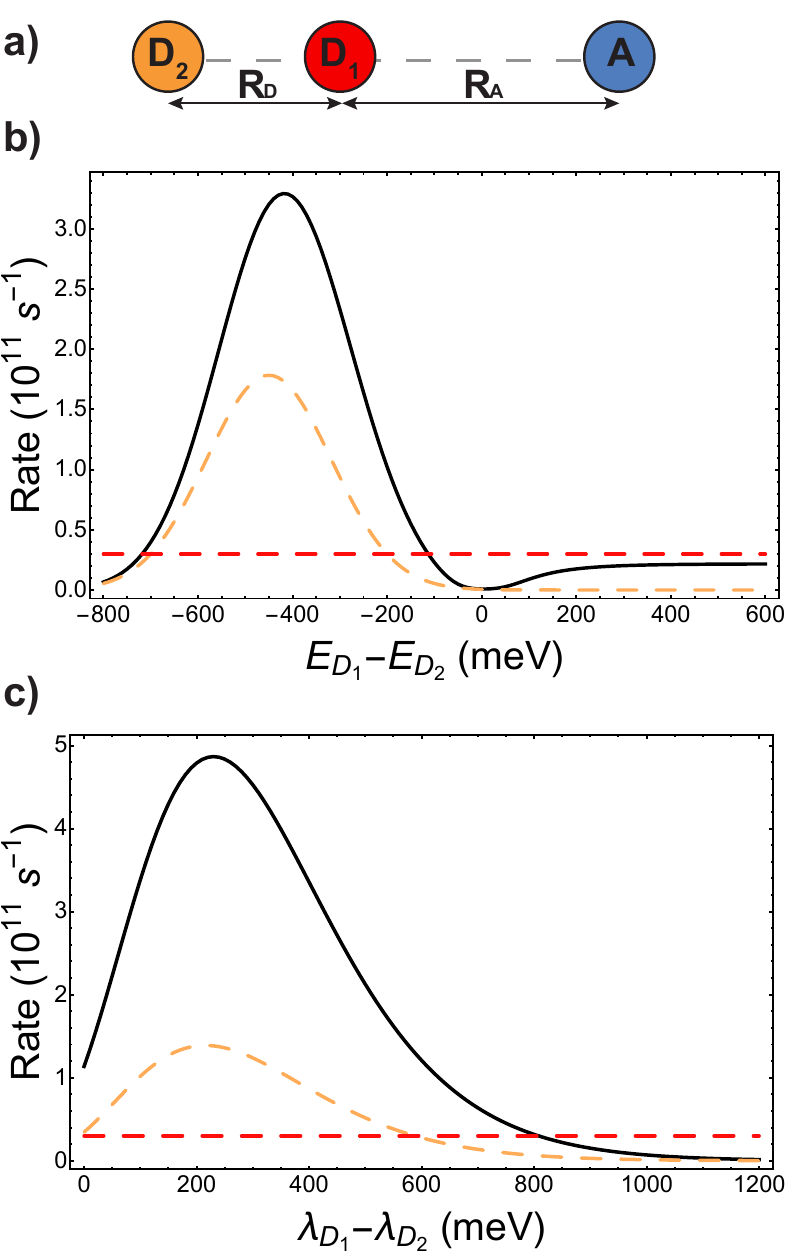}
		\caption{\label{fenergy} Tuning energy offsets and reorganisation energies can enhance charge-transfer rates beyond what is possible with either donor site alone.
			a) Two donors and one acceptor in a collinear geometry, with different colours (orange/red) indicating inequivalent donors.
			b) Energetic detuning: the charge transfer rate from the aggregate to the acceptor (solid line) is compared to the rate if only $D_1$ (dashed red) or $D_2$ (dashed orange) were present, as a function of the energy difference between $D_1$ and $D_2$. Even with the effects of supertransfer removed (the aggregate rate is shown divided by the electronic supertransfer enhancement of 1.42), energetic tuning can make the aggregate transfer faster than would be possible with either donor alone. In particular, the presence of $D_2$, which itself is weakly coupled to $A$, can enhance the transfer rate above the rate from $D_1$ alone.
			c) Reorganisation energies: plot as in b), but the rates are shown as a function of the difference in reorganisation energies between $D_1$ and $D_2$. Here as well, adding $D_2$ with a favourable reorganisation energy can enhance the rate above what is possible with either donor alone.
			Calculation parameters: $V_{D_1D_2}=\SI{-37}{\milli\electronvolt}$, $V_{D_1A}=\SI{18}{\milli\electronvolt}$, $V_{D_2A}=\SI{2.5}{\milli\electronvolt}$, $\lambda_A=\SI{200}{\milli\electronvolt}$, $\lambda_{D_1}=\SI{150}{\milli\electronvolt}$, $E_{D_1}=\SI{700}{\milli\electronvolt}$, $E_A=\SI{0}{\milli\electronvolt}$, $k_BT=\SI{25}{\milli\electronvolt}$. In addition, b) uses $\lambda_{D_2}=\SI{150}{\milli\electronvolt}$ and c) has $E_{D_2}=\SI{600}{\milli\electronvolt}$.
		}
	\end{figure}

	The presence of different processes affecting the nuclear factor means that delocalisation can have a complicated effect on the charge-transfer rate, even apart from supertransfer. Critical to the rate is the relative size of $\Delta E_{\alpha \beta}$ and $\lambda_{\alpha \beta}$, because of the rate's exponential sensitivity to $(\Delta E_{\alpha \beta}+\lambda_{\alpha \beta})^2$. The different effects are illustrated with another example, shown in Fig.~\ref{fenergy}a, where acceptor $A$ is strongly coupled to donor $D_1$, whose site energy and reorganisation energy are such that the transfer from $D_1$ to $A$ is very slow ($-\Delta E_{D_1A}\gg\lambda_{D_1A}$). Another donor $D_2$ is then introduced, but is weakly coupled to $A$ due to its distance. A naive application of classical MT might suggest that, because $D_2$ hardly interacts with $A$, it would serve to only steal charge density from $D_1$, reducing the already slow transfer rate. Generalised MT, however, shows that it is possible to choose the energy and reorganisation energy of $D_2$, as well as its coupling to $D_1$, so that a coherent superposition between $D_1$ and $D_2$ will enhance the total transfer rate above what is possible with \emph{either} $D_1$ or $D_2$ alone. This is true even if supertransfer is neglected, as shown in Figs.~\ref{fenergy}b and~c. Indeed, for two donors, supertransfer can enhance the rate by at most a factor of two, while there is no limit to how much the nuclear factor can be enhanced by judiciously tuning $\Delta E_{\alpha \beta}$ and $\lambda_{\alpha \beta}$ to minimise $(\Delta E_{\alpha \beta}+\lambda_{\alpha \beta})^2$. This result shows that even if an unfavourable donor must be used in a donor-acceptor system (for whatever reason), another donor can be added to tune the nuclear term's contribution to the charge transfer rate.

	Our results also extend gMT to treat bridge-mediated charge transfer, showing that the usual equations still apply when considering delocalised aggregates. Indeed, including the effects of bridge-mediated charge transfer on gMT does not qualitatively change the effects of supertransfer and energetic tuning, except that the coherent effects depend on the geometry of the donor aggregate with respect to the first bridge molecule, and the acceptor aggregate with respect to the last. In particular, the results shown in Figs.~\ref{fangle} and~\ref{fenergy} would remain unchanged if the couplings were mediated by a bridge.

	\section{Conclusion} \label{conclusion}
	
	The theory presented in this work is the first description of charge transfer between delocalised molecular aggregates. Therefore, we anticipate that it will have broad applications in fields where charge transfer and electronic coherence intersect, including organic photovoltaics, photosynthesis, and inorganic complexes.
	
	The major prediction of gMT is that delocalisation within an aggregate can significantly affect charge transfer rates through two mechanisms: supertransfer and nuclear tuning. The first is a consequence of the constructive interferences of charge-transfer pathways, while the latter is the ability of a charge-transfer rate to be modified by adjusting effective energy levels and reorganisation energies by delocalising electronic states over different molecules.
	
	Both of these predictions are suited to being tested experimentally. The simplest approach would be to construct covalently linked donors and acceptors in geometries that approximate those in Figs.~\ref{fangle} and~\ref{fenergy}. Tuning the couplings and energy levels through chemical modification would permit the adjustment of the parameters relevant for gMT, allowing the theory to be tested.
	
	In this work, we restricted ourselves to deriving the delocalised generalisation of the simplest Marcus-theory formula. We are confident that many of the subsequent advances that have occurred in charge-transfer theory can also be incorporated as extensions to gMT. Indeed, our derivation is more general than the final result, and some of the approximations needed to derive an MT-like equation (e.g., high temperature, slow environmental modes) can be omitted and more general intermediate results used directly (e.g., Equations~\ref{iediag}--\ref{ieAfinal}). Although it is not clear whether a simple, closed-form expression could be derived, a number of improvements to gMT can be envisaged, including adiabatic charge transfer, quantum-mechanical vibrational corrections~\cite{Ulstrup1975}, coherent multistep charge transfer~\cite{Kuznetsov2000a}, shared intra-aggregate environmental modes~\cite{Renger2002}, and off-diagonal system-environment couplings. Inspiration could also be taken from advances in MC-FRET to obtain generalisations able to treat system-environment entanglement or other parameter regimes outside the approximations used here~\cite{Ma2015,Ma2015:2,Moix2015}.
	
	\section{Appendix} \label{appendix}
	
	Here we give the full derivation of Eqs.~\ref{iedint}--\ref{ieaint} from Eq.~\ref{ebegin}, indexing the sum with $k''$ for future convenience:	
	\begin{equation}\label{ekredfield}
	k_{D\rightarrow A}(t)=\frac{d}{dt}\mathrm{Tr}_E\sum_{k''}\bra{A_k''}\rho(t)\ket{A_k''}.
	\end{equation}
	
	Since the inter-aggregate coupling $H_C$ is weak compared to all other terms in $H$, we take it as a perturbation. Taking $H_0=H-H_C$, and using tildes to denote the interaction picture, we write $\tilde{H}_C(t)=e^{iH_0t/\hbar}H_Ce^{-iH_0t/\hbar}$ and express $\dot{\tilde{\rho}}(t)$ to second order in perturbation theory:
	\begin{equation}
	\dot{\tilde{\rho}}(t)\approx -\frac{1}{\hbar^2}\int_0^t d\tau\, \mathrm{Tr}_E\left[ \tilde{H}_C(t),\left[ \tilde{H}_C(\tau),\rho(0)\right]\right],
	\end{equation}
	where $[\cdot,\cdot]$ is the commutator, and $\mathrm{Tr}_E$ is the trace over the environment degrees of freedom. Substituting into Eq.~\ref{ekredfield},
	\begin{align}\label{einteraction}
	k_{D\rightarrow A}(t)=&-\frac{1}{\hbar^2}\sum_{k''}\Bra{A_k''}\int_0^t d\tau\nonumber\\& \mathrm{Tr}_E\left[ \tilde{H}_C(t),\left[ \tilde{H}_C(\tau),\rho(0)\right]\right]\Ket{A_k''}\\
	=&-\frac{1}{\hbar^2}\sum_{k''}\Bra{A_k''}\int_0^t d\tau\, \mathrm{Tr}_E \big(\tilde{H}_C(t)\tilde{H}_C(\tau)\rho(0)\nonumber\\&+\rho(0)\tilde{H}_C(\tau)\tilde{H}_C(t)-\tilde{H}_C(\tau)\rho(0)\tilde{H}_C(t)\nonumber\\&-\tilde{H}_C(t)\rho(0)\tilde{H}_C(\tau)\big)\Ket{A_k''}.
	\end{align}
	Since the charge is initially on the donor aggregate, $\rho(0)\ket{A_k''}=\bra{A_k''}\rho(0)=0$, the first two terms vanish, giving
	\begin{align}\label{combined}
	k_{D\rightarrow A}(t)=&\frac{1}{\hbar^2}\sum_{k''}2\mathrm{Re}\Bra{A_k''}\int_0^t d\tau\nonumber\\& \mathrm{Tr}_E \big(\tilde{H}_C(\tau)\rho(0)\tilde{H}_C(t)\big)\Ket{A_k''} \\
	=&\sum_{j,j'}\sum_{k,k',k''}\frac{V_{jk}V_{j' k'}}{\hbar^2}\cdot 2\mathrm{Re} \int_0^t d\tau\nonumber\\&\mathrm{Tr}_E \big(\Bra{A_{k''}}e^{iH_0\tau/\hbar}\Ket{A_{k'}} \Bra{D_{j'}}e^{-iH_0\tau/\hbar}\rho(0)\nonumber \\& e^{iH_0t/\hbar}\Ket{D_j}\Bra{A_k}e^{-iH_0t/\hbar} \Ket{A_{k''}} \big).
	\end{align}
	Using the cyclic property of the trace gives
	\begin{multline}\label{eexpandedform}
	k_{D\rightarrow A}(t)=\sum_{j,j'}\sum_{k,k'}\frac{V_{jk}V_{j' k'}}{\hbar^2}\cdot 2\mathrm{Re}
	\int_0^t d\tau\,\\\mathrm{Tr}_E\big(
	\Bra{A_k}e^{-iH_0(t-\tau)/\hbar}\Ket{A_{k'}}\\
	\Bra{D_{j'}}e^{-iH_0\tau/\hbar}\rho(0)e^{iH_0t/\hbar}\Ket{D_j}
	\big).
	\end{multline}
	Defining $\tau'=t-\tau$, we can write
	\begin{multline}
	k_{D\rightarrow A}(t)=\sum_{j,j'}\sum_{k,k'}\frac{V_{jk}V_{j' k'}}{\hbar^2}\cdot \int_{-t}^td\tau'\\\mathrm{Tr}_E\big(\Bra{A_k}e^{-iH_0\tau'/\hbar}\Ket{A_{k'}}
	\Bra{D_{j'}}e^{iH_0(\tau'-t)/\hbar}\rho(0)\\e^{-iH_0(\tau'-t)/\hbar}e^{iH_0\tau'/\hbar}\Ket{D_j}\big).\label{eexpandedform2}
	\end{multline}
	
	To simplify further, we consider the term $e^{iH_0(\tau'-t)/\hbar}\rho(0)e^{-iH_0(\tau'-t)/\hbar}$, which describes the time-evolution of the donor aggregate (because $H_0$ induces no donor-acceptor transitions). Because the aggregate-environment coupling is much stronger than the inter-aggregate coupling, the donor aggregate will thermalise with the environment on timescales much shorter than the charge-transfer timescale. Therefore, for times $t$ much longer than the donor thermalisation time (but much shorter than the charge-transfer time), we can consider the long-time limit,
	\begin{equation}
	\lim\limits_{t\rightarrow\infty} e^{iH_0(\tau'-t)/\hbar}\rho(0)e^{-iH_0(\tau'-t)/\hbar}=\rho_{\mathrm{th}}.
	\end{equation}
	where for a large, weakly coupled environment, the state $\rho_{\mathrm{th}}=\rho_D\otimes\rho_A$, of donor and acceptor aggregates independently thermalised with their own environments, is independent of $\rho(0)$. In this limit, we may also extend the limits of integration in Equation~\ref{eexpandedform2} to infinity to give a time-independent rate:
	\begin{multline}\label{eexpandedform3}
	k_{D\rightarrow A}=\sum_{j,j'}\sum_{k,k'}\frac{V_{jk}V_{j' k'}}{\hbar^2}\cdot \int_{-\infty}^\infty d\tau'\mathrm{Tr}_E\big(\\\Bra{A_k}e^{-iH_0\tau'/\hbar}\Ket{A_{k'}}
	\Bra{D_{j'}}\rho_{\mathrm{th}}e^{iH_0\tau'/\hbar}\Ket{D_j}\big).
	\end{multline}
	
	Writing $H_D=H_D^0+H_{DE}+H_{E_D}$ and $H_A=H_A^0+H_{AE}+H_{E_A}$ and using Plancherel's theorem, we can  rewrite Eq.~\ref{eexpandedform3} as
	\begin{align}\label{eratesite}
	k_{D\rightarrow A} =& \sum_{j,j'}\sum_{k,k'}\frac{V_{jk}V_{j' k'}}{2\pi\hbar^2}\int_{-\infty}^{\infty}d\omega\, \mathcal{D}_D^{jj'}(\omega)\mathcal{A}_A^{kk'}(\omega),\\
	\label{edisassociation}
	\mathcal{D}_D^{jj'}(\omega) =&  \int_{-\infty}^{\infty}dt\,e^{-i\omega t}\mathrm{Tr}_{E_D}\big(\nonumber\\& e^{-iH_{E_D}t/\hbar}
	\Bra{D_{j'}}e^{iH_D t/\hbar}\rho_D\Ket{D_{j}}\big),\\
	\label{eassociation}
	\mathcal{A}_A^{kk'}(\omega) =& \int_{-\infty}^{\infty}dt\,e^{i\omega t}\mathrm{Tr}_{E_A}\big(\nonumber\\& e^{iH_{E_A}t/\hbar}
	\Bra{A_{k}}e^{-iH_A t/\hbar}\Ket{A_{k'}}\rho_A\big),
	\end{align}
	where we have renamed $\tau'$ to $t$.
	Changing to the aggregate basis, Eq.~\ref{eratesite} becomes
	\begin{align}\label{eoffdiag}
	k_{D\rightarrow A}=&\sum_{\alpha,\beta}\sum_{\alpha',\beta'}\frac{V_{\alpha\beta}V_{\alpha'\beta'}}{2\pi\hbar^2}\int_{-\infty}^\infty d\omega\,\mathcal{D}_D^{\alpha\alpha'}(\omega)\mathcal{A}_A^{\beta\beta'}(\omega),\\
	\mathcal{D}_D^{\alpha\alpha'}(\omega) =&  \int_{-\infty}^{\infty}dt\,e^{-i\omega t}\mathrm{Tr}_{E_D}\big(\nonumber\\& e^{-iH_{E_D}t/\hbar}
	\Bra{D_{\alpha'}}e^{iH_D t/\hbar}\rho_D\Ket{D_{\alpha}}\big),\\
	\mathcal{A}_A^{\beta\beta'}(\omega) =& \int_{-\infty}^{\infty}dt\,e^{i\omega t}\mathrm{Tr}_{E_A}\big(\nonumber\\& e^{iH_{E_A}t/\hbar}
	\Bra{A_{\beta}}e^{-iH_A t/\hbar}\Ket{A_{\beta'}}\rho_A\big),
	\end{align}
	
	Eq.~\ref{eoffdiag} reduces to Eq.~\ref{iediag} if $\mathcal{D}_D^{\alpha\alpha'}$ and $\mathcal{A}_A^{\beta\beta'}$ can be assumed to be diagonal in the aggregate basis. In general, this is not the case, because $H_{DE}$ and $H_{AE}$ do not commute with $H_D^0$ and $H_A^0$ respectively. However, it is an appropriate approximation in the limit, assumed here, of weak system-environment coupling, where the environment does not significantly perturb the thermal equilibrium of the system. The same approximation was considered and discussed in detail in the context of MC-FRET~\cite{Jang2004,Ma2015}, where it can be used to reduce the excitonic analogue of Eq.~\ref{eoffdiag} to a diagonal version. Of course, Eq.~\ref{eoffdiag} can be used directly, at the cost of intuitive parallels with MT being obscured.
	
	Eqs.~\ref{ieDfinal}--\ref{ieAfinal} can be evaluated in the particular case of a thermalised environment of independent harmonic oscillators to yield Eqs.~\ref{iedint}--\ref{ieaint}. Assuming that $\mathcal{D}_D^{\alpha\alpha'}$ and $\mathcal{A}_A^{\beta\beta'}$ are diagonal is equivalent to assuming that the electronic Hamiltonians commute with the environmental ones, meaning that $\exp(iH_Dt/\hbar)=\exp(iH_D^0 t/\hbar)\exp(i(H_{DE}+H_{E_D})t/\hbar)$, so that Eq.~\ref{ieDfinal} becomes
	\begin{multline}
	\mathcal{D}_D^{\alpha\alpha}(\omega)=
	\int_{-\infty}^{\infty}dt\,e^{-i\omega t}e^{i E_{\alpha}t/\hbar}\rho_{\alpha\alpha} \mathrm{Tr}_{E_D}\big(\\e^{-iH_{E_D}t/\hbar}\Bra{D_{\alpha}}e^{i(H_{DE}+H_{E_D}) t/\hbar}\Ket{D_{\alpha}}\rho_{E_D}\big).\label{eq:polaron-step1}
	\end{multline}
	The Hamiltonian $H_{DE}+H_{E_D}$ can be diagonalised using the polaron transformation, which describes the displacement of the environment oscillators by the presence of a charge:
	\begin{equation}
	\bra{D_\alpha}e^{i(H_{DE}+H_{E_D}) t/\hbar}\ket{D_\alpha}=S_{\alpha}^\dagger e^{iH_{E_D} t/\hbar} S_{\alpha},
	\end{equation}
	where $S_{\alpha}=\exp\big(\sum_\xi g_{\alpha\xi}(b_{\xi}-b_{\xi}^\dagger)\big)$. Using this fact in Eq.~\ref{eq:polaron-step1} gives
	\begin{multline}
	\mathcal{D}_D^{\alpha\alpha}(\omega)=
	\int_{-\infty}^{\infty}dt\,e^{-i\omega t}e^{i E_{\alpha}t/\hbar}\rho_{\alpha\alpha} \mathrm{Tr}_{E_D}\big(\\e^{-i{H_{E_D}}t/\hbar}  S_{{\alpha}}^\dagger e^{i{H_{E_D}}t/\hbar} S_{{\alpha}} \rho_{E_D} \big).\label{eddisp}
	\end{multline}
	In Eq.~\ref{eddisp}, the contributions of different aggregate eigenstates are explicitly uncoupled, meaning that the equation takes, for a particular $\alpha$, the same form that occurs in the derivation of ordinary, single-site MT. Therefore, the trace can be evaluated for a harmonic environment using standard techniques (e.g., section 6.8.1 of ref.~\cite{MayKuhn}), giving Eq.~\ref{iedint}.
	
	\section{Acknowledgements}
	We thank Tom Stace and the UQ Node EQuS Theory Group for helpful comments on the manuscript. We were supported by the Westpac Bicentennial Foundation through a Westpac Research Fellowship and by the Australian Research Council through a Discovery Early Career Researcher Award (DE140100433) and  the Centre of Excellence for Engineered Quantum Systems (CE110001013).
	

\begin{thebibliography}{34}%
\makeatletter
\providecommand \@ifxundefined [1]{%
 \@ifx{#1\undefined}
}%
\providecommand \@ifnum [1]{%
 \ifnum #1\expandafter \@firstoftwo
 \else \expandafter \@secondoftwo
 \fi
}%
\providecommand \@ifx [1]{%
 \ifx #1\expandafter \@firstoftwo
 \else \expandafter \@secondoftwo
 \fi
}%
\providecommand \natexlab [1]{#1}%
\providecommand \enquote  [1]{``#1''}%
\providecommand \bibnamefont  [1]{#1}%
\providecommand \bibfnamefont [1]{#1}%
\providecommand \citenamefont [1]{#1}%
\providecommand \href@noop [0]{\@secondoftwo}%
\providecommand \href [0]{\begingroup \@sanitize@url \@href}%
\providecommand \@href[1]{\@@startlink{#1}\@@href}%
\providecommand \@@href[1]{\endgroup#1\@@endlink}%
\providecommand \@sanitize@url [0]{\catcode `\\12\catcode `\$12\catcode
  `\&12\catcode `\#12\catcode `\^12\catcode `\_12\catcode `\%12\relax}%
\providecommand \@@startlink[1]{}%
\providecommand \@@endlink[0]{}%
\providecommand \url  [0]{\begingroup\@sanitize@url \@url }%
\providecommand \@url [1]{\endgroup\@href {#1}{\urlprefix }}%
\providecommand \urlprefix  [0]{URL }%
\providecommand \Eprint [0]{\href }%
\providecommand \doibase [0]{http://dx.doi.org/}%
\providecommand \selectlanguage [0]{\@gobble}%
\providecommand \bibinfo  [0]{\@secondoftwo}%
\providecommand \bibfield  [0]{\@secondoftwo}%
\providecommand \translation [1]{[#1]}%
\providecommand \BibitemOpen [0]{}%
\providecommand \bibitemStop [0]{}%
\providecommand \bibitemNoStop [0]{.\EOS\space}%
\providecommand \EOS [0]{\spacefactor3000\relax}%
\providecommand \BibitemShut  [1]{\csname bibitem#1\endcsname}%
\let\auto@bib@innerbib\@empty
\bibitem [{\citenamefont {Marcus}\ and\ \citenamefont
  {Sutin}(1985)}]{Marcus1985a}%
  \BibitemOpen
  \bibfield  {author} {\bibinfo {author} {\bibfnamefont {R.}~\bibnamefont
  {Marcus}}\ and\ \bibinfo {author} {\bibfnamefont {N.}~\bibnamefont {Sutin}},\
  }\href {\doibase 10.1016/0304-4173(85)90014-X} {\bibfield  {journal}
  {\bibinfo  {journal} {Biochim. Biophys. Acta -- Rev. Bioenerg.}\ }\textbf
  {\bibinfo {volume} {811}},\ \bibinfo {pages} {265} (\bibinfo {year}
  {1985})}\BibitemShut {NoStop}%
\bibitem [{\citenamefont {Barbara}\ \emph {et~al.}(1996)\citenamefont
  {Barbara}, \citenamefont {Meyer},\ and\ \citenamefont
  {Ratner}}]{Barbara1996a}%
  \BibitemOpen
  \bibfield  {author} {\bibinfo {author} {\bibfnamefont {P.~F.}\ \bibnamefont
  {Barbara}}, \bibinfo {author} {\bibfnamefont {T.~J.}\ \bibnamefont {Meyer}},
  \ and\ \bibinfo {author} {\bibfnamefont {M.~A.}\ \bibnamefont {Ratner}},\
  }\href {\doibase 10.1021/jp9605663} {\bibfield  {journal} {\bibinfo
  {journal} {J. Phys. Chem.}\ }\textbf {\bibinfo {volume} {100}},\ \bibinfo
  {pages} {13148} (\bibinfo {year} {1996})}\BibitemShut {NoStop}%
\bibitem [{\citenamefont {May}\ and\ \citenamefont {Kühn}(2011)}]{MayKuhn}%
  \BibitemOpen
  \bibfield  {author} {\bibinfo {author} {\bibfnamefont {V.}~\bibnamefont
  {May}}\ and\ \bibinfo {author} {\bibfnamefont {O.}~\bibnamefont {Kühn}},\
  }\href@noop {} {\emph {\bibinfo {title} {Charge and Energy Transfer in
  Molecular Systems}}},\ \bibinfo {edition} {3rd}\ ed.\ (\bibinfo  {publisher}
  {Wiley-VCH},\ \bibinfo {address} {Weinheim, Germany},\ \bibinfo {year}
  {2011})\BibitemShut {NoStop}%
\bibitem [{\citenamefont {Jortner}\ and\ \citenamefont
  {Bixon}(1999)}]{jortner1999electron}%
  \BibitemOpen
  \bibfield  {author} {\bibinfo {author} {\bibfnamefont {J.}~\bibnamefont
  {Jortner}}\ and\ \bibinfo {author} {\bibfnamefont {M.}~\bibnamefont
  {Bixon}},\ }\href@noop {} {\emph {\bibinfo {title} {Adv. Chem. Phys.}}},\
  Vol.\ \bibinfo {volume} {106--107}\ (\bibinfo {year} {1999})\BibitemShut
  {NoStop}%
\bibitem [{\citenamefont {Few}\ \emph {et~al.}(2015)\citenamefont {Few},
  \citenamefont {Frost},\ and\ \citenamefont {Nelson}}]{Few2014}%
  \BibitemOpen
  \bibfield  {author} {\bibinfo {author} {\bibfnamefont {S.}~\bibnamefont
  {Few}}, \bibinfo {author} {\bibfnamefont {J.~M.}\ \bibnamefont {Frost}}, \
  and\ \bibinfo {author} {\bibfnamefont {J.}~\bibnamefont {Nelson}},\ }\href
  {\doibase 10.1039/C4CP03663H} {\bibfield  {journal} {\bibinfo  {journal}
  {Phys. Chem. Chem. Phys.}\ }\textbf {\bibinfo {volume} {17}},\ \bibinfo
  {pages} {2311} (\bibinfo {year} {2015})}\BibitemShut {NoStop}%
\bibitem [{\citenamefont {K{\"o}hler}\ and\ \citenamefont
  {B{\"a}ssler}(2015)}]{kohler2015electronic}%
  \BibitemOpen
  \bibfield  {author} {\bibinfo {author} {\bibfnamefont {A.}~\bibnamefont
  {K{\"o}hler}}\ and\ \bibinfo {author} {\bibfnamefont {H.}~\bibnamefont
  {B{\"a}ssler}},\ }\href@noop {} {\emph {\bibinfo {title} {Electronic
  Processes in Organic Semiconductors: An Introduction}}}\ (\bibinfo
  {publisher} {John Wiley \& Sons},\ \bibinfo {year} {2015})\BibitemShut
  {NoStop}%
\bibitem [{\citenamefont {Allen}\ and\ \citenamefont
  {Williams}(1998)}]{Allen1998}%
  \BibitemOpen
  \bibfield  {author} {\bibinfo {author} {\bibfnamefont {J.~P.}\ \bibnamefont
  {Allen}}\ and\ \bibinfo {author} {\bibfnamefont {J.~C.}\ \bibnamefont
  {Williams}},\ }\href {\doibase 10.1104/pp.125.1.33} {\bibfield  {journal}
  {\bibinfo  {journal} {FEBS Lett.}\ }\textbf {\bibinfo {volume} {438}},\
  \bibinfo {pages} {5} (\bibinfo {year} {1998})}\BibitemShut {NoStop}%
\bibitem [{\citenamefont {Blankenship}(2014)}]{Blankenship}%
  \BibitemOpen
  \bibfield  {author} {\bibinfo {author} {\bibfnamefont {R.}~\bibnamefont
  {Blankenship}},\ }\href@noop {} {\emph {\bibinfo {title} {Molecular
  Mechanisms of Photosynthesis}}},\ \bibinfo {edition} {2nd}\ ed.\ (\bibinfo
  {publisher} {Wiley Blackwell},\ \bibinfo {address} {Chichester, UK},\
  \bibinfo {year} {2014})\BibitemShut {NoStop}%
\bibitem [{\citenamefont {D'Alessandro}\ and\ \citenamefont
  {Keene}(2006)}]{DAlessandro2006a}%
  \BibitemOpen
  \bibfield  {author} {\bibinfo {author} {\bibfnamefont {D.~M.}\ \bibnamefont
  {D'Alessandro}}\ and\ \bibinfo {author} {\bibfnamefont {F.~R.}\ \bibnamefont
  {Keene}},\ }\href {\doibase 10.1021/cr050010o} {\bibfield  {journal}
  {\bibinfo  {journal} {Chem. Rev.}\ }\textbf {\bibinfo {volume} {106}},\
  \bibinfo {pages} {2270} (\bibinfo {year} {2006})}\BibitemShut {NoStop}%
\bibitem [{\citenamefont {D'Alessandro}\ \emph {et~al.}(2011)\citenamefont
  {D'Alessandro}, \citenamefont {Kanga},\ and\ \citenamefont
  {Caddy}}]{DAlessandro2011}%
  \BibitemOpen
  \bibfield  {author} {\bibinfo {author} {\bibfnamefont {D.~M.}\ \bibnamefont
  {D'Alessandro}}, \bibinfo {author} {\bibfnamefont {J.~R.~R.}\ \bibnamefont
  {Kanga}}, \ and\ \bibinfo {author} {\bibfnamefont {J.~S.}\ \bibnamefont
  {Caddy}},\ }\href {\doibase 10.1071/CH11039} {\bibfield  {journal} {\bibinfo
  {journal} {Aust. J. Chem.}\ }\textbf {\bibinfo {volume} {64}},\ \bibinfo
  {pages} {718} (\bibinfo {year} {2011})}\BibitemShut {NoStop}%
\bibitem [{\citenamefont {Hush}(1958)}]{Hush1958}%
  \BibitemOpen
  \bibfield  {author} {\bibinfo {author} {\bibfnamefont {N.~S.}\ \bibnamefont
  {Hush}},\ }\href {\doibase 10.1063/1.1744305} {\bibfield  {journal} {\bibinfo
   {journal} {J. Chem. Phys.}\ }\textbf {\bibinfo {volume} {28}},\ \bibinfo
  {pages} {962} (\bibinfo {year} {1958})}\BibitemShut {NoStop}%
\bibitem [{\citenamefont {Hush}(1961)}]{Hush1960}%
  \BibitemOpen
  \bibfield  {author} {\bibinfo {author} {\bibfnamefont {N.~S.}\ \bibnamefont
  {Hush}},\ }\href {\doibase 10.1039/tf9615700557} {\bibfield  {journal}
  {\bibinfo  {journal} {Trans. Faraday Soc.}\ }\textbf {\bibinfo {volume}
  {57}},\ \bibinfo {pages} {557} (\bibinfo {year} {1961})}\BibitemShut
  {NoStop}%
\bibitem [{\citenamefont {Cave}\ and\ \citenamefont
  {Newton}(1996)}]{Cave:1996ep}%
  \BibitemOpen
  \bibfield  {author} {\bibinfo {author} {\bibfnamefont {R.~J.}\ \bibnamefont
  {Cave}}\ and\ \bibinfo {author} {\bibfnamefont {M.~D.}\ \bibnamefont
  {Newton}},\ }\href@noop {} {\bibfield  {journal} {\bibinfo  {journal} {Chem.
  Phys. Lett.}\ }\textbf {\bibinfo {volume} {249}},\ \bibinfo {pages} {15}
  (\bibinfo {year} {1996})}\BibitemShut {NoStop}%
\bibitem [{\citenamefont {Matyushov}\ and\ \citenamefont
  {Voth}(2000)}]{Matyushov2000d}%
  \BibitemOpen
  \bibfield  {author} {\bibinfo {author} {\bibfnamefont {D.~V.}\ \bibnamefont
  {Matyushov}}\ and\ \bibinfo {author} {\bibfnamefont {G.~A.}\ \bibnamefont
  {Voth}},\ }\href {\doibase 10.1021/jp993885d} {\bibfield  {journal} {\bibinfo
   {journal} {J. Phys. Chem. A}\ }\textbf {\bibinfo {volume} {104}},\ \bibinfo
  {pages} {6470} (\bibinfo {year} {2000})}\BibitemShut {NoStop}%
\bibitem [{\citenamefont {Subotnik}\ \emph {et~al.}(2008)\citenamefont
  {Subotnik}, \citenamefont {Yeganeh}, \citenamefont {Cave},\ and\
  \citenamefont {Ratner}}]{Subotnik:2008fp}%
  \BibitemOpen
  \bibfield  {author} {\bibinfo {author} {\bibfnamefont {J.~E.}\ \bibnamefont
  {Subotnik}}, \bibinfo {author} {\bibfnamefont {S.}~\bibnamefont {Yeganeh}},
  \bibinfo {author} {\bibfnamefont {R.~J.}\ \bibnamefont {Cave}}, \ and\
  \bibinfo {author} {\bibfnamefont {M.~A.}\ \bibnamefont {Ratner}},\
  }\href@noop {} {\bibfield  {journal} {\bibinfo  {journal} {J. Chem. Phys.}\
  }\textbf {\bibinfo {volume} {129}},\ \bibinfo {pages} {244101} (\bibinfo
  {year} {2008})}\BibitemShut {NoStop}%
\bibitem [{\citenamefont {Hush}(1967)}]{hush1967intervalence}%
  \BibitemOpen
  \bibfield  {author} {\bibinfo {author} {\bibfnamefont {N.~S.}\ \bibnamefont
  {Hush}},\ }\href@noop {} {\bibfield  {journal} {\bibinfo  {journal} {Prog.
  Inorg. Chem.}\ }\textbf {\bibinfo {volume} {8}},\ \bibinfo {pages} {12}
  (\bibinfo {year} {1967})}\BibitemShut {NoStop}%
\bibitem [{\citenamefont {Marcus}(1956)}]{Marcus1956a}%
  \BibitemOpen
  \bibfield  {author} {\bibinfo {author} {\bibfnamefont {R.~A.}\ \bibnamefont
  {Marcus}},\ }\href {\doibase 10.1063/1.1742723} {\bibfield  {journal}
  {\bibinfo  {journal} {J. Chem. Phys.}\ }\textbf {\bibinfo {volume} {24}},\
  \bibinfo {pages} {966} (\bibinfo {year} {1956})}\BibitemShut {NoStop}%
\bibitem [{\citenamefont {Levich}\ and\ \citenamefont
  {Dogonadze}(1959)}]{levich1959theory}%
  \BibitemOpen
  \bibfield  {author} {\bibinfo {author} {\bibfnamefont {V.}~\bibnamefont
  {Levich}}\ and\ \bibinfo {author} {\bibfnamefont {R.}~\bibnamefont
  {Dogonadze}},\ }\href@noop {} {\bibfield  {journal} {\bibinfo  {journal}
  {Dokl. Akad. Nauk. SSSR}\ }\textbf {\bibinfo {volume} {124}},\ \bibinfo
  {pages} {123} (\bibinfo {year} {1959})}\BibitemShut {NoStop}%
\bibitem [{\citenamefont {Sumi}(1999)}]{Sumi1999}%
  \BibitemOpen
  \bibfield  {author} {\bibinfo {author} {\bibfnamefont {H.}~\bibnamefont
  {Sumi}},\ }\href {\doibase 10.1021/jp983477u} {\bibfield  {journal} {\bibinfo
   {journal} {J. Phys. Chem. B}\ }\textbf {\bibinfo {volume} {103}},\ \bibinfo
  {pages} {252} (\bibinfo {year} {1999})}\BibitemShut {NoStop}%
\bibitem [{\citenamefont {Sumi}(2001)}]{Sumi2001a}%
  \BibitemOpen
  \bibfield  {author} {\bibinfo {author} {\bibfnamefont {H.}~\bibnamefont
  {Sumi}},\ }\href {\doibase 10.1002/tcr.10004} {\bibfield  {journal} {\bibinfo
   {journal} {Chem. Rec.}\ }\textbf {\bibinfo {volume} {1}},\ \bibinfo {pages}
  {480} (\bibinfo {year} {2001})}\BibitemShut {NoStop}%
\bibitem [{\citenamefont {Scholes}\ \emph {et~al.}(2001)\citenamefont
  {Scholes}, \citenamefont {Jordanides},\ and\ \citenamefont
  {Fleming}}]{Scholes2001}%
  \BibitemOpen
  \bibfield  {author} {\bibinfo {author} {\bibfnamefont {G.}~\bibnamefont
  {Scholes}}, \bibinfo {author} {\bibfnamefont {X.}~\bibnamefont {Jordanides}},
  \ and\ \bibinfo {author} {\bibfnamefont {G.}~\bibnamefont {Fleming}},\ }\href
  {\doibase 10.1021/jp003571m} {\bibfield  {journal} {\bibinfo  {journal} {J.
  Phys. Chem. B}\ }\textbf {\bibinfo {volume} {105}},\ \bibinfo {pages} {1640}
  (\bibinfo {year} {2001})}\BibitemShut {NoStop}%
\bibitem [{\citenamefont {Baghbanzadeh}\ and\ \citenamefont
  {Kassal}(2016{\natexlab{a}})}]{Baghbanzadeh2016}%
  \BibitemOpen
  \bibfield  {author} {\bibinfo {author} {\bibfnamefont {S.}~\bibnamefont
  {Baghbanzadeh}}\ and\ \bibinfo {author} {\bibfnamefont {I.}~\bibnamefont
  {Kassal}},\ }\href {\doibase 10.1021/acs.jpclett.6b01779} {\bibfield
  {journal} {\bibinfo  {journal} {J. Phys. Chem. Lett.}\ }\textbf {\bibinfo
  {volume} {7}},\ \bibinfo {pages} {3804} (\bibinfo {year}
  {2016}{\natexlab{a}})},\ \Eprint {http://arxiv.org/abs/1604.05482}
  {1604.05482} \BibitemShut {NoStop}%
\bibitem [{\citenamefont {Baghbanzadeh}\ and\ \citenamefont
  {Kassal}(2016{\natexlab{b}})}]{Baghbanzadeh2016a}%
  \BibitemOpen
  \bibfield  {author} {\bibinfo {author} {\bibfnamefont {S.}~\bibnamefont
  {Baghbanzadeh}}\ and\ \bibinfo {author} {\bibfnamefont {I.}~\bibnamefont
  {Kassal}},\ }\href {\doibase 10.1039/c6cp00104a} {\bibfield  {journal}
  {\bibinfo  {journal} {Phys. Chem. Chem. Phys.}\ }\textbf {\bibinfo {volume}
  {18}},\ \bibinfo {pages} {7459} (\bibinfo {year}
  {2016}{\natexlab{b}})}\BibitemShut {NoStop}%
\bibitem [{\citenamefont {Jang}\ \emph {et~al.}(2004)\citenamefont {Jang},
  \citenamefont {Newton},\ and\ \citenamefont {Silbey}}]{Jang2004}%
  \BibitemOpen
  \bibfield  {author} {\bibinfo {author} {\bibfnamefont {S.}~\bibnamefont
  {Jang}}, \bibinfo {author} {\bibfnamefont {M.}~\bibnamefont {Newton}}, \ and\
  \bibinfo {author} {\bibfnamefont {R.}~\bibnamefont {Silbey}},\ }\href
  {\doibase 10.1103/PhysRevLett.92.218301} {\bibfield  {journal} {\bibinfo
  {journal} {Phys. Rev. Lett.}\ }\textbf {\bibinfo {volume} {92}},\ \bibinfo
  {pages} {218301} (\bibinfo {year} {2004})}\BibitemShut {NoStop}%
\bibitem [{\citenamefont {Ma}\ and\ \citenamefont {Cao}(2015)}]{Ma2015}%
  \BibitemOpen
  \bibfield  {author} {\bibinfo {author} {\bibfnamefont {J.}~\bibnamefont
  {Ma}}\ and\ \bibinfo {author} {\bibfnamefont {J.}~\bibnamefont {Cao}},\
  }\href {\doibase 10.1063/1.4908599} {\bibfield  {journal} {\bibinfo
  {journal} {J. Chem. Phys.}\ }\textbf {\bibinfo {volume} {142}},\ \bibinfo
  {pages} {094106} (\bibinfo {year} {2015})}\BibitemShut {NoStop}%
\bibitem [{\citenamefont {Cleary}\ and\ \citenamefont
  {Cao}(2013)}]{Cleary2013}%
  \BibitemOpen
  \bibfield  {author} {\bibinfo {author} {\bibfnamefont {L.}~\bibnamefont
  {Cleary}}\ and\ \bibinfo {author} {\bibfnamefont {J.}~\bibnamefont {Cao}},\
  }\href {\doibase 10.1088/1367-2630/15/12/125030} {\bibfield  {journal}
  {\bibinfo  {journal} {New J. Phys.}\ }\textbf {\bibinfo {volume} {15}},\
  \bibinfo {pages} {125030} (\bibinfo {year} {2013})}\BibitemShut {NoStop}%
\bibitem [{\citenamefont {Nitzan}(2006)}]{Nitzan}%
  \BibitemOpen
  \bibfield  {author} {\bibinfo {author} {\bibfnamefont {A.}~\bibnamefont
  {Nitzan}},\ }\href@noop {} {\emph {\bibinfo {title} {Chemical Dynamics in
  Condensed Phases}}}\ (\bibinfo  {publisher} {Oxford University Press},\
  \bibinfo {address} {Oxford, UK},\ \bibinfo {year} {2006})\BibitemShut
  {NoStop}%
\bibitem [{\citenamefont {Jang}\ and\ \citenamefont {Cheng}(2013)}]{Jang2013a}%
  \BibitemOpen
  \bibfield  {author} {\bibinfo {author} {\bibfnamefont {S.}~\bibnamefont
  {Jang}}\ and\ \bibinfo {author} {\bibfnamefont {Y.-C.}\ \bibnamefont
  {Cheng}},\ }\href {\doibase 10.1002/wcms.1111} {\bibfield  {journal}
  {\bibinfo  {journal} {WIREs Comput. Mol. Sci.}\ }\textbf {\bibinfo {volume}
  {3}},\ \bibinfo {pages} {84} (\bibinfo {year} {2013})}\BibitemShut {NoStop}%
\bibitem [{\citenamefont {Fruchtman}\ \emph {et~al.}(2016)\citenamefont
  {Fruchtman}, \citenamefont {G{\'{o}}mez-Bombarelli}, \citenamefont {Lovett},\
  and\ \citenamefont {Gauger}}]{Fruchtman2016}%
  \BibitemOpen
  \bibfield  {author} {\bibinfo {author} {\bibfnamefont {A.}~\bibnamefont
  {Fruchtman}}, \bibinfo {author} {\bibfnamefont {R.}~\bibnamefont
  {G{\'{o}}mez-Bombarelli}}, \bibinfo {author} {\bibfnamefont {B.~W.}\
  \bibnamefont {Lovett}}, \ and\ \bibinfo {author} {\bibfnamefont {E.~M.}\
  \bibnamefont {Gauger}},\ }\href {\doibase 10.1103/PhysRevLett.117.203603}
  {\bibfield  {journal} {\bibinfo  {journal} {Phys. Rev. Lett.}\ }\textbf
  {\bibinfo {volume} {117}},\ \bibinfo {pages} {203603} (\bibinfo {year}
  {2016})},\ \Eprint {http://arxiv.org/abs/1511.06302} {arXiv:1511.06302}
  \BibitemShut {NoStop}%
\bibitem [{\citenamefont {Ulstrup}\ and\ \citenamefont
  {Jortner}(1975)}]{Ulstrup1975}%
  \BibitemOpen
  \bibfield  {author} {\bibinfo {author} {\bibfnamefont {J.}~\bibnamefont
  {Ulstrup}}\ and\ \bibinfo {author} {\bibfnamefont {J.}~\bibnamefont
  {Jortner}},\ }\href {\doibase 10.1063/1.431152} {\bibfield  {journal}
  {\bibinfo  {journal} {J. Chem. Phys.}\ }\textbf {\bibinfo {volume} {63}},\
  \bibinfo {pages} {4358} (\bibinfo {year} {1975})}\BibitemShut {NoStop}%
\bibitem [{\citenamefont {Kuznetsov}\ and\ \citenamefont
  {Ulstrup}(2000)}]{Kuznetsov2000a}%
  \BibitemOpen
  \bibfield  {author} {\bibinfo {author} {\bibfnamefont {A.}~\bibnamefont
  {Kuznetsov}}\ and\ \bibinfo {author} {\bibfnamefont {J.}~\bibnamefont
  {Ulstrup}},\ }\href {\doibase 10.1016/S0013-4686(00)00336-4} {\bibfield
  {journal} {\bibinfo  {journal} {Electrochim. Acta}\ }\textbf {\bibinfo
  {volume} {45}},\ \bibinfo {pages} {2339} (\bibinfo {year}
  {2000})}\BibitemShut {NoStop}%
\bibitem [{\citenamefont {Renger}\ and\ \citenamefont
  {Marcus}(2002)}]{Renger2002}%
  \BibitemOpen
  \bibfield  {author} {\bibinfo {author} {\bibfnamefont {T.}~\bibnamefont
  {Renger}}\ and\ \bibinfo {author} {\bibfnamefont {R.~A.}\ \bibnamefont
  {Marcus}},\ }\href {\doibase 10.1063/1.1470200} {\bibfield  {journal}
  {\bibinfo  {journal} {J. Chem. Phys.}\ }\textbf {\bibinfo {volume} {116}},\
  \bibinfo {pages} {9997} (\bibinfo {year} {2002})}\BibitemShut {NoStop}%
\bibitem [{\citenamefont {Ma}\ \emph {et~al.}(2015)\citenamefont {Ma},
  \citenamefont {Moix},\ and\ \citenamefont {Cao}}]{Ma2015:2}%
  \BibitemOpen
  \bibfield  {author} {\bibinfo {author} {\bibfnamefont {J.}~\bibnamefont
  {Ma}}, \bibinfo {author} {\bibfnamefont {J.}~\bibnamefont {Moix}}, \ and\
  \bibinfo {author} {\bibfnamefont {J.}~\bibnamefont {Cao}},\ }\href {\doibase
  10.1063/1.4908600} {\bibfield  {journal} {\bibinfo  {journal} {J. Chem.
  Phys.}\ }\textbf {\bibinfo {volume} {142}},\ \bibinfo {pages} {094107}
  (\bibinfo {year} {2015})},\ \Eprint {http://arxiv.org/abs/1405.4771}
  {1405.4771} \BibitemShut {NoStop}%
\bibitem [{\citenamefont {Moix}\ \emph {et~al.}(2015)\citenamefont {Moix},
  \citenamefont {Ma},\ and\ \citenamefont {Cao}}]{Moix2015}%
  \BibitemOpen
  \bibfield  {author} {\bibinfo {author} {\bibfnamefont {J.~M.}\ \bibnamefont
  {Moix}}, \bibinfo {author} {\bibfnamefont {J.}~\bibnamefont {Ma}}, \ and\
  \bibinfo {author} {\bibfnamefont {J.}~\bibnamefont {Cao}},\ }\href {\doibase
  10.1063/1.4908601} {\bibfield  {journal} {\bibinfo  {journal} {J. Chem.
  Phys.}\ }\textbf {\bibinfo {volume} {142}},\ \bibinfo {pages} {094108}
  (\bibinfo {year} {2015})}\BibitemShut {NoStop}%
\end{thebibliography}

%

\end{document}